\documentclass[a4paper,11pt]{article}
\pdfoutput=1 

\usepackage{mathrsfs}
\usepackage{mathtools}
\usepackage{slashbox}
\usepackage{jcappub}
\usepackage{amsmath,amssymb}
\usepackage{graphicx}
\usepackage{latexsym}
\usepackage{xspace}
\usepackage{color}
\usepackage{hyperref}
\usepackage{bm}
\usepackage{relsize}
\usepackage{tabularx}
\usepackage{multirow}
\usepackage{amssymb}
\usepackage[table]{xcolor}
\usepackage{colortbl}
\usepackage{pdflscape}
\usepackage{bigints}
\usepackage{graphics}
\usepackage{adjustbox}

\makeatletter
\gdef\@fpheader{}
\g@addto@macro\bfseries{\boldmath}
\makeatother

\DeclarePairedDelimiterX\Basics[1](){ #1}

\DeclarePairedDelimiterX{\infdivx}[2]{(}{)}{%
  #1\;\delimsize\|\;#2%
}

\newcommand{\dkl}{D_{\rm KL}}

\newcommand{\dd}{\mathrm{d}}
\newcommand{\ee}{e}

\newcommand{\boldsigma}{\boldsymbol{\sigma}}
\newcommand{\boldz}{\boldsymbol{z}}

\newcommand{\sss}[1]{{\scriptscriptstyle{#1}}}

\newcommand{\uPl}{\mathrm{Pl}}

\newcommand{\uS}{\mathrm{S}}

\newcommand{\usssS}{\sss{\uS}}

\newcommand{\usssPl}{\sss{\uPl}}

\newcommand{\nS}{n_\usssS}

\newcommand{\alphaS}{\alpha_\usssS}

\newcommand{\calM}{\mathcal{M}}

\newcommand{\like}{{\cal L}}

\newcommand{\bmuf}{\boldsymbol{\mu}_{{}_{\rm F}}}

\newcommand{\muf}{\mu_{{}_{{\rm F}}}}

\newcommand{\boldx}{\boldsymbol{x}}
\newcommand{\boldy}{\boldsymbol{y}}

\newcommand{\Mp}{M_\usssPl}

\newcommand{\deci}{\mathscr{D}}

\newcommand{\efold}{$e$-fold}

\newcommand{\beq}{\begin{equation}}
\newcommand{\eeq}{\end{equation}}
\newcommand{\bea}{\begin{eqnarray}}
\newcommand{\eea}{\end{eqnarray}}

\newlength{\wsingfig}
\newlength{\wdblefig}
\newlength{\wquadfig}
\newlength{\wtriplefig}
\newcommand{\barpi}{\bar{\pi}}
\newcommand{\barp}{\bar{p}}
\newcommand{\barlike}{\bar{{\cal L}}}
\newcommand{\hatpi}{\hat{\pi}}
\newcommand{\hatp}{\hat{p}}
\newcommand{\hatlike}{\hat{{\cal L}}}

\newcommand{\Eq}[1]{Eq.~(\ref{#1})}

\newcommand{\Fig}[1]{Fig.~{\ref{#1}}}

\newcommand{\Ref}[1]{Ref.~{\cite{#1}}}

\newcommand{\Sec}[1]{Sec.~\ref{#1}}

\subheader{}

\title{The decisive future of inflation}

\author[a]{Robert J. Hardwick,}
\author[a,b]{Vincent Vennin,}
\author[a]{David Wands}

\affiliation[a]{Institute of Cosmology \& Gravitation, University of Portsmouth, Dennis Sciama Building, Burnaby Road, Portsmouth, PO1 3FX, United Kingdom}
\affiliation[b]{Laboratoire Astroparticule et Cosmologie, Universit\'{e} Denis Diderot Paris 7,
10 rue Alice Domon et L\'{e}onie Duquet, 75013 Paris, France}

\emailAdd{robert.hardwick@port.ac.uk}
\emailAdd{vincent.vennin@port.ac.uk}
\emailAdd{david.wands@port.ac.uk}

\date{today}

\begin{document}
\sloppy

\abstract{How much more will we learn about single-field inflationary models in the future? We address this question in the context of Bayesian design and information theory. We develop a novel method to compute the expected utility of deciding between models and apply it to a set of futuristic measurements. This necessarily requires one to evaluate the Bayesian evidence many thousands of times over, which is numerically challenging. We show how this can be done using a number of simplifying assumptions and discuss their validity. We also modify the form of the expected utility, as previously introduced in the literature in different contexts, in order to partition each possible future into either the rejection of models at the level of the maximum likelihood or the decision between models using Bayesian model comparison.  We then quantify the ability of future experiments to constrain the reheating temperature and the scalar running. Our approach allows us to discuss possible strategies for maximising information from future cosmological surveys. In particular, our conclusions suggest that, in the context of inflationary model selection, a decrease in the measurement uncertainty of the scalar spectral index would be more decisive than a decrease in the uncertainty in the tensor-to-scalar ratio. We have incorporated our approach into a publicly available python class, \href{https://sites.google.com/view/foxicode}{\texttt{foxi}}, that can be readily applied to any survey optimisation problem.}



\maketitle
\section{Introduction}
\label{sec:intro}
The recent \emph{Planck} collaboration results~\cite{Ade:2013sjv,Adam:2015rua,Ade:2015lrj} marked a significant milestone in model selection for inflation using data from the Cosmic Microwave Background (CMB). In the case of single-field models, the decreased upper bound on the tensor-to-scalar ratio combined with a red-tilted spectral index lead the analysis to mostly favour inflationary potentials with a plateau~\cite{Easther:2011yq,Planck:2013jfk,Martin:2013nzq,Ade:2015lrj,Vennin:2015eaa}. Additionally, multi-field inflation has also recently begun to be rigorously statistically analysed, e.g. in the context of curvaton models~\cite{Hardwick:2015tma,Vennin:2015egh,dePutter:2016trg}.

Despite the significant reduction in the number of observationally viable models, it has become abundantly clear that there are still quite a number of models that satisfy the \emph{Planck} constraints, especially those classed in the plateau category of potential. This dissatisfying state of affairs is only mitigated by the potential for other future surveys to augment the current constraints such as CMB Stage-4~\cite{Abazajian:2016yjj}, LiteBIRD~\cite{litebird} and  COrE~\cite{DiValentino:2016foa,Finelli:2016cyd}. Despite the promise of further observations, the future of inflationary model selection is still tremendously unclear. In the face of an uncertain future, we seek to answer the following question: To what extent can one be certain of a future survey being capable of deciding between models, or within the space of many models? The answer is probabilistic and clearly dependent not only on the particular model choice, but also on the current constraints made by the \emph{Planck} collaboration. Since a decision must be made, the natural framework to answer this question uses Bayesian probability.

\begin{figure}[t]
\begin{center}
\includegraphics[width=0.65\textwidth]{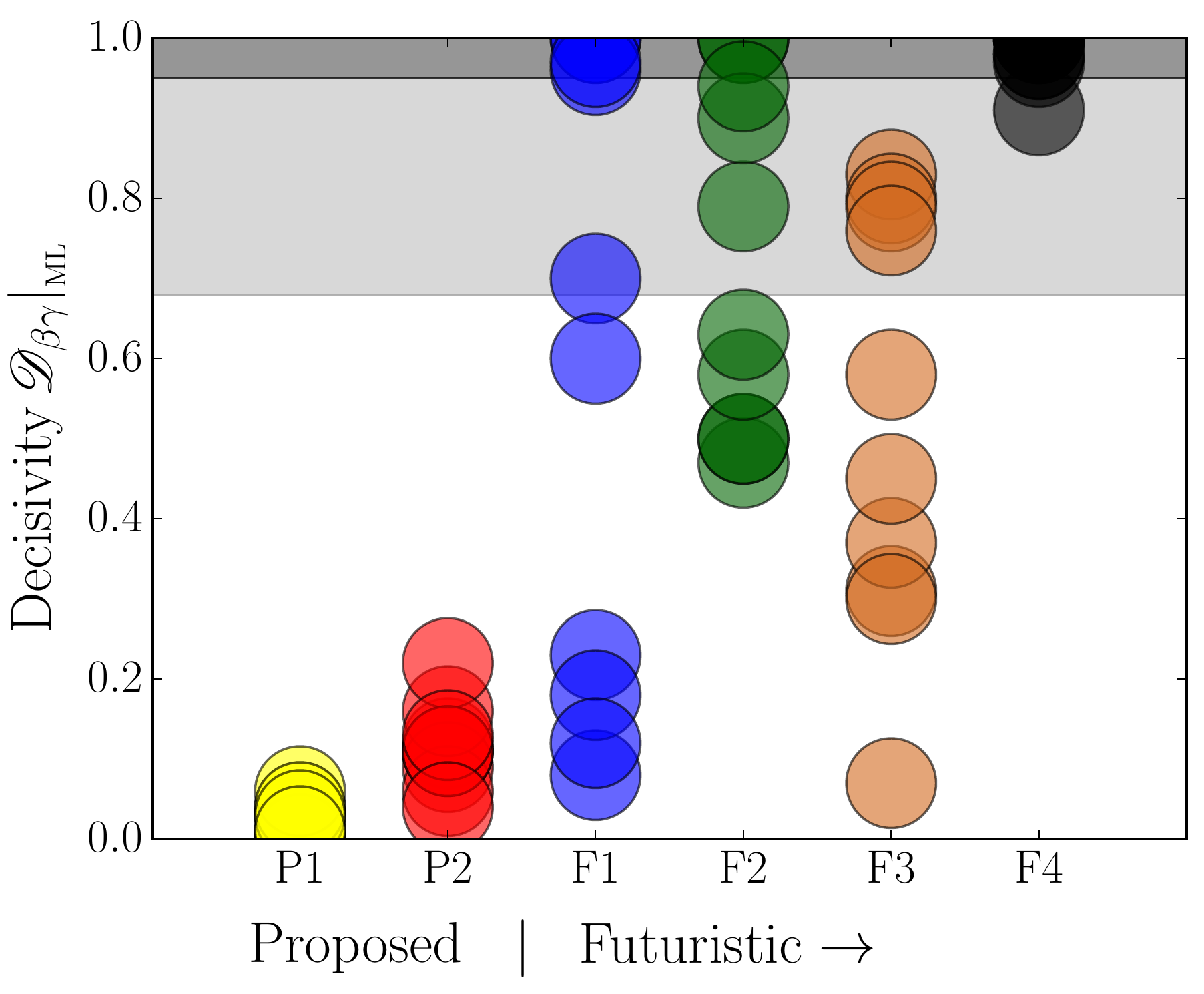}
\caption{~\label{fig:allmodels_deci} A scatter plot of each model pair score in the decisivity utility $\deci_{\beta \gamma}\vert_{{}_{\rm ML}}$ (computed using the maximum-likelihood average, see \Eq{eq:ml-average}) using the Bayes factors of each of the possible pairs of models for each futuristic survey, and the $5$ representative single-field models used in this work. We have assumed a logarithmic prior over $\epsilon_1$ (\Eq{eq:predictive-priors-logeps1}) and a flat prior over $(\epsilon_2,\epsilon_3)$ --- see also the discussion in \Sec{sec:formalism}. The light and dark grey rectangles correspond to $\deci_{\beta \gamma}\vert_{{}_{\rm ML}}=0.68$ and $0.95$ i.e. to situations where the probability to rule out one model against the other is $68\%$ and $95\%$, respectively. The colours and labels on the horizontal axis correspond to the measurement configurations of Table \ref{tab:future-widths}. }
\end{center}
\end{figure}

It seems clear that there are many interesting unanswered questions one can pose relating to the predictive probabilities of future survey performance. In this work, we will restrict ourselves to focus on using a futuristic set of measurement widths to compute our defined expected utilities for model distinguishability. Therefore, the specific question we pose for this paper is as follows: \emph{How much more do we stand to learn about single-field inflationary models given a forecast set of future measurement widths over the slow-roll parameters?} To this end, we set up six classes of survey over the space of slow-roll parameters $(\epsilon_1 , \epsilon_2, \epsilon_3 )$, defined using the Hubble parameter $H$ and its derivatives with respect to the number of \efold{s} ($N\equiv \ln a$, where $a$ is the scale factor) like so
\begin{equation}
\epsilon_0 \equiv \frac{1}{H} \,, \qquad \epsilon_{i+1} \equiv \frac{\dd {\ln}\vert \epsilon_i\vert}{\dd N}\,,
\end{equation}
where the corresponding choices of measurement widths (for $i=1,2,3$) of each fictitious experiment are defined in Table \ref{tab:future-widths}. Our expectation will be a clear trend between decreasing measurement widths and an improvement in the score from our utility functions, e.g. as can be seen from \Fig{fig:allmodels_deci}, where we have plotted the quantity $\deci_{\beta \gamma} \vert_{{}_{\rm ML}}$ --- defined as a score of decisive merit between models in later chapters --- against our mock surveys.

We acknowledge that the broad question we seek to answer in this paper has been approached, to some degree, at various angles by Refs.~\cite{Martin:2014rqa,Hardwick:2015tma,Finelli:2016cyd} (though no work yet appears to apply this to CMB experiments and models of inflation). In each case, the authors target a slightly different problem with specific surveys in mind. Further to this, we note that some of the quantities we will later define (such as $\deci$) have already been introduced in similar works for Dark Energy models~\cite{2011MNRAS.414.2337T}, likelihood parameter inference for \emph{Planck}~\cite{Trotta:2007hy} and to classify the cosmic web in~\cite{Leclercq:2016fyj} --- yet the formalism will be extended and improved in this work to properly quantify the ability of future surveys to distinguish between models of inflation.
\begin{table}
\centering
\begin{tabular}{|l*{7}{|c}|}\hline
\multicolumn{2}{|c}{Reference} & \multicolumn{3}{|c|}{Measurements} & \multicolumn{2}{c|}{$\langle \dkl \rangle$ }   \\\hline
Name & Colour &$\sigma^1 (\epsilon_1 )$ & $\sigma^2 (\epsilon_2 )$ & $\sigma^3 (\epsilon_3 )$  & $\pi (\bmuf \vert \, \epsilon_1)$ & $\pi (\bmuf \vert \, \log \epsilon_1)$
\\\hline\hline
\cellcolor{yellow!55} & Proposed 1 (P1) & $10^{-3}$ & $10^{-2}$ & $10^{-1}$ & 5.6 $\pm$ 0.3 & 0.6 $\pm$ 1.1 \\\hline
\cellcolor{red!55} & Proposed 2 (P2) & $10^{-4}$ & $10^{-2}$ & $10^{-2}$ & 9.9 $\pm$ 0.5 & 2.1 $\pm$ 2.1 \\\hline
\cellcolor{blue!35} & Futuristic 1 (F1) & $10^{-5}$ & $10^{-2}$ & $10^{-2}$ & $>11.4$ & 2.5 $\pm$ 2.5 \\\hline
\cellcolor{green!55} & Futuristic 2 (F2) & $10^{-4}$ & $10^{-3}$ & $10^{-2}$ &  $>11.4$ & 3.5 $\pm$ 2.7 \\\hline
\cellcolor{brown!90} & Futuristic 3 (F3) & $10^{-4}$ & $10^{-2}$ & $10^{-3}$ &  $>11.4$ & 4.1 $\pm$ 2.2 \\\hline
\cellcolor{black!65} & Futuristic 4 (F4) & $10^{-5}$ & $10^{-3}$ & $10^{-3}$ &  $>11.4$ & 5.7 $\pm$ 3.0 \\\hline
\end{tabular}
\caption{~\label{tab:future-widths} Measurement accuracy (in terms of the $1$-$\sigma$ error bars on the first three slow-roll parameters) and expected Kullback-Leibler divergence (information gain) between the prior and posterior distributions over the slow-roll parameters for the future toy surveys studied in this work. The first two are set with similar characteristics to potential surveys in the near future and are denoted P1 and P2 (CMB Stage-4 and COrE/LiteBIRD, respectively, where `P' stands for `Proposed'). In addition, we have exceeded these forecasts with our Futuristic categories 1-4 (F1-4) to indicate various (possibly absolute) limits. We direct the reader to \Sec{sec:formalism} for the discussion that motivates the $\epsilon_1$ flat ($\pi (\bmuf \vert \, \epsilon_1 )$) and the $\epsilon_1$ logarithmic ($\pi (\bmuf \vert \log \epsilon_1)$) priors. The $\langle \dkl\rangle > 11.4$ values using a flat prior over $\epsilon_1$ exceed a numerical threshold associated to the integral computation of \Eq{eq:calcDKL}.}
\end{table}

In this paper we will outline a simple method to compute any expected utility for a future survey given a previous set of measurements on the same variables from an independent survey (which, in our case, shall always be the \emph{Planck} 2015 constraints). In \Sec{sec:formalism} we outline in detail our definition of the utility functions to be used throughout this work, as well as introducing some new methods of computation --- including our outline of the new \href{https://sites.google.com/view/foxicode}{\texttt{foxi}} algorithm.

The \href{https://sites.google.com/view/foxicode}{\texttt{foxi}} (Futuristic Observations and their eXpected Information) package is a general-purpose, publicly available, python class for use on any forecasting problem. It outputs \LaTeX \, compile-able tables and has a variety of plotting options. One can fork the code and other details through the website: \href{https://sites.google.com/view/foxicode}{https://sites.google.com/view/foxicode}. We have also included some robustness checks and a brief summary of the computational methods used by the algorithm in appendix~\ref{sec:foxi-computation}.

\begin{figure}[t]
\begin{center}
\includegraphics[width=0.6\textwidth]{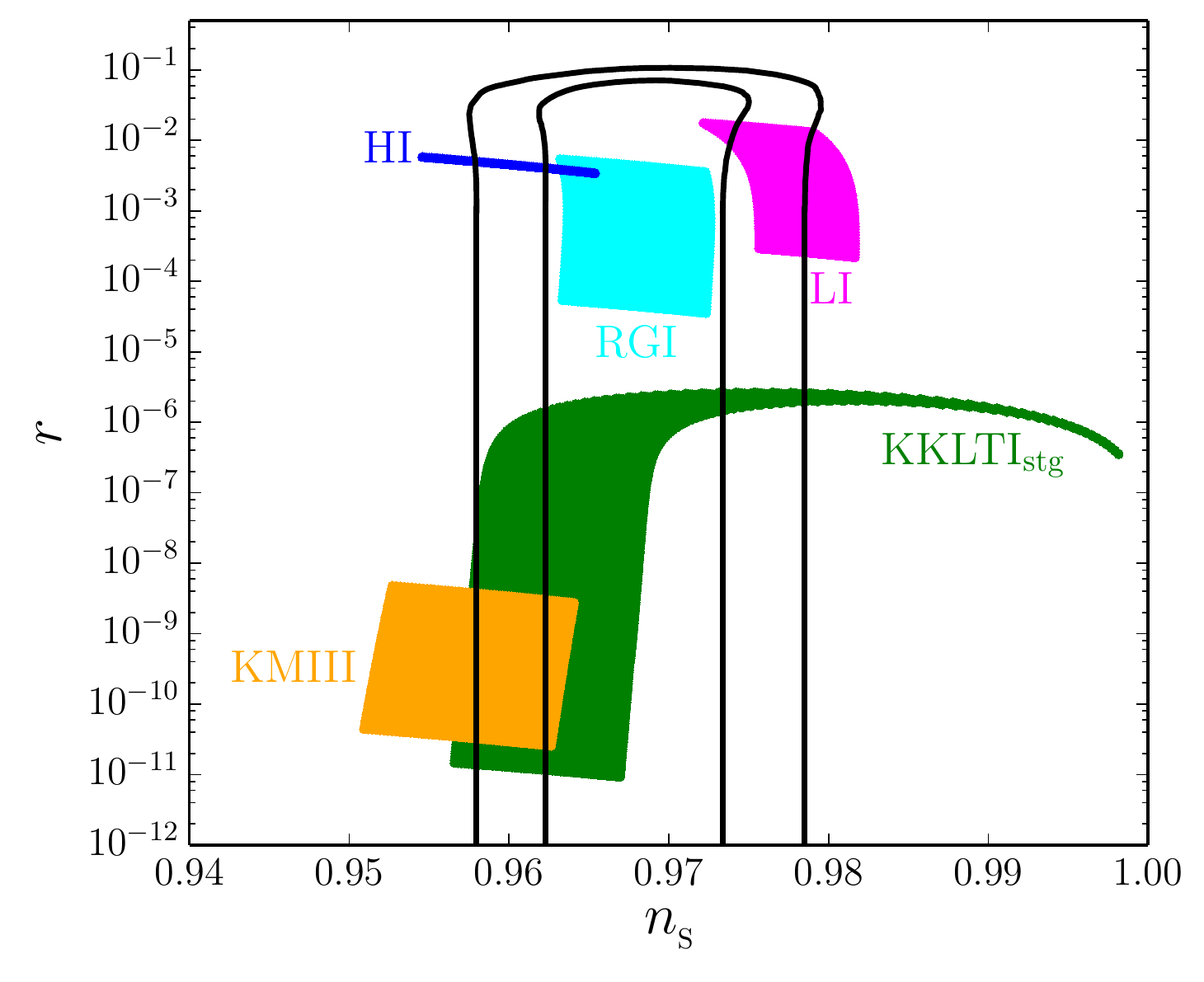}
\caption{~\label{fig:nsr-plot} An $(\nS , r)$-plot of the available parameter space to each of the models used in this work, where the solid black contours are the 68\% and 95\% limits currently imposed by the \emph{Planck} 2015 data~\cite{Ade:2015lrj}. $\nS$ on the horizontal axis is the scalar spectral index and $r$ on the vertical axis is the tensor-to-scalar ratio. }
\end{center}
\end{figure}

Since literally hundreds of single-field models have been proposed in the literature~\cite{Martin:2013tda}, including all of them in our analysis would be numerically too expensive. In order to infer results that are representative of the full model set one must therefore choose a variety of models that fill e.g. the $(\nS ,r)$ diagram, using
\begin{align}
\nS - 1 &\equiv \left. \frac{\dd \ln {\cal P}_{\zeta}}{\dd \ln k} \right\vert_{k_*} \simeq 1 - 2\epsilon_1 - \epsilon_2\,, \\
\ r &\equiv \left. \frac{{\cal P}_{h}}{{\cal P}_{\zeta}} \right\vert_{k_*} \simeq 16\epsilon_1\,, \label{eq:tens-scalar-ratio}
\end{align}
where ${\cal P}_\zeta$ and ${\cal P}_h$ are the power spectra for scalar and tensor perturbations, respectively, $k_*=0.05\,{\rm Mpc}^{-1}$ denotes the pivot scale and the last equalities in both expressions are valid only for single-field models to leading order in slow roll. In appendix~\ref{sec:models} we list the 5 representative single-field models --- employed in the \href{http://cp3.irmp.ucl.ac.be/~ringeval/aspic.html}{\texttt{ASPIC}} library~\cite{Martin:2013tda,aspic}: Higgs Inflation (HI), K\"{a}hler Moduli Inflation II (KMIII), Kachru-Kallosh-Linde-Trivedi Inflation (${\rm KKLTI}_{\rm stg}$), Loop Inflation (${\rm LI}_{\alpha >0}$) and Radion Gauge Inflation (RGI) --- that we have chosen, neglecting many reasonable alternatives for the sake of brevity and capturing the essential information about the competition between models. Though no favouritism for these 5 is intended in this paper,\footnote{\href{https://sites.google.com/view/foxicode}{\texttt{foxi}} copes relatively well with the inclusion of many models, though the number of model pairs to analyse scales with the Binomial coefficient $\frac{N!}{(N-2)!2!}$, where $N$ is the number of models. Already with $N=5$, we note that $10$ model pairs must be considered.} as they are merely representative of the explored parameter space shown by our representation of each prior volume over the $(\nS , r)$-plane in \Fig{fig:nsr-plot}, we nonetheless have provided very brief introduction for each (which includes both their potentials and priors on their parameters) in appendix~\ref{sec:models}.

Our results can be found in \Sec{sec:results}, where we employ a comprehensive suite of expected utilities to analyse the future of model selection for inflation.
We have additionally included a small section (\Sec{sec:reheating}) on the interesting possibility of using our framework to examine the future prospects of inferring the reheating temperature in the example of the HI model as well as a computation of the probability in the future that each of the various survey configurations will be able to exceed a $2$-$\sigma$ detection of the running of the scalar spectral index $\alphaS$ in \Sec{sec:measuring-alphaS} (with a preliminary calculation in appendix~\ref{sec:alphaS-calculation}). Both of these short examples are intended to give an impression of the possible scope of usage for our code \href{https://sites.google.com/view/foxicode}{\texttt{foxi}} with a model-focused question in mind. Finally, in \Sec{sec:concl} we present our conclusions.

\section{Formalism}
\label{sec:formalism}
\subsection{Probability measures primer}
Due to the fact that all of the models of inflation considered here are slow-roll models, there exists a general parameterisation of the power spectrum (which we observe) that includes $n$ slow-roll parameters ${\cal P}_{\zeta}={\cal P}_{\zeta} (\epsilon_1,\epsilon_2,\epsilon_3,\dots , \epsilon_n )$ that is sufficient to constrain their observational characteristics once the amplitude has been measured and fixed. The current data, using \emph{Planck} CMB measurements~\cite{Ade:2013sjv,Adam:2015rua,Ade:2015lrj}, limits our capabilities to constrain up to essentially $n=3$ slow-roll parameters~\cite{Ringeval:2013lea,Ade:2013sjv,Adam:2015rua,Ade:2015lrj}. Even though future surveys may in principle be able to constrain parameters further up the slow-roll hierarchy, e.g. $\epsilon_4$, they will first need to constrain $\epsilon_3$ at the level that is consistent with slow roll, which we find to be difficult even for the most futuristic of our toy surveys considered here (see \Sec{sec:measuring-alphaS}). Hence, though all of the formalism in this work can be applied to any $n$-dimensional parameter spaces, we shall consider here only the space of slow-roll parameters $(\epsilon_1,\epsilon_2,\epsilon_3)$ as a first example. This space will subsequently be equipped with three distinct probability measures.

\subsection*{The posterior given the current data}
Hereafter, the fiducial point vector $\bmuf$ spans the real $n$-dimensional parameter space of central points for future measurements. This, naturally, has a probability measure associated to it which is derived from the current observations over each separate direction in the space. We can therefore define the integral measure over the domain of $\bmuf$ (such as will be used in \Eq{eq:exu}) as the posterior distribution of current data $p \, (\bmuf \vert {\cal D}_{\rm cur} ) \, \dd \bmuf$. There is a subtlety in obtaining $p \,(\bmuf |{\cal D}_{\rm cur} )$, that is revealed through Bayes' rule
\begin{equation} \label{eq:post-pred}
\ p \,(\bmuf |{\cal D}_{\rm cur} ) \propto   \pi_{{}_{\cal I}}  (\bmuf ) \, \like\,({\cal D}_{\rm cur} | \bmuf ) \,,
\end{equation}
which includes the prior information $\pi_{{}_{\cal I}}  (\bmuf )$ over the space of $\bmuf$, the former containing some initial information ${\cal I}$ about the sampling space\footnote{This may be constructed from either subjective theoretical prejudice, invariant volumes under group transformations of the parameter space (see e.g. the Haar measure) or information provided from the likelihood itself --- e.g. the `Jeffreys prior', which is $\propto \sqrt{\det {\cal F}_{ij}}$, where ${\cal F}_{ij}$ is the Fisher information matrix.}.

Within the specific choice of parameterisation $(\epsilon_1,\epsilon_2,\epsilon_3)$, throughout this work we will make two choices of prior where $\muf^2 \in [0,0.09] \,,   \,\, \muf^3 \in [-0.2,0.2]$ and
\begin{align} \label{eq:predictive-priors-flateps1}
\pi \, (\bmuf | \, \epsilon_1  ) \propto {\rm const.} \,, \quad &{\rm where} \quad \muf^1 \in [10^{-4},10^{-2}] \,, \\
\pi \, (\bmuf | \log \epsilon_1 ) \propto \frac{1}{\muf^1} \,, \,\,\, \quad &{\rm where} \quad \log (\muf^1) \in [-13,-1] \,, \label{eq:predictive-priors-logeps1}
\end{align}
corresponding to either flat, or, flat in all dimensions except a log prior over the component $\muf^1$ i.e. the first slow-roll parameter $\epsilon_1$, respectively. By setting the hard prior limits in \Eq{eq:predictive-priors-logeps1}, we have artificially chosen the lower bound on $\epsilon_1 = 10^{-13}$, which seems reasonable when none of the models we study here are capable of lower values than this and, in the absence of an absolute lower fundamental limit\footnote{
We restrict $\epsilon_1\geq 10^{-13}$, otherwise we would need to include second-order effects in perturbation theory~\cite{Martineau:2007dj}. In addition, this lower bound encompasses the predictions from all of our chosen model priors.} on $r$, that limit is also placed so as to not overweight too much of the prior volume on very low values which will likely never be detectable. The upper limit on $\epsilon_1$ and the bounds on both $\epsilon_2$ and $\epsilon_3$ are set by slow-roll consistency.

To give an indication of the volume of permitted $\bmuf$ points used in this work, the $\pi \, (\bmuf | \log \epsilon_1 )$ prior has been used in \Fig{fig:nsr-plot} to display the 68\% and 95\% contour limits (in solid black) for the current \emph{Planck} 2015 posterior marginalised over the $(\nS , r)$-plane. 

\subsection*{The prior from each model}

We define $\boldx$ as a real $n$-dimensional vector over the same observables represented by $\bmuf$ (hence, for this work it is over $(\epsilon_1,\epsilon_2,\epsilon_3)$). To generate a model prior $\bar{\pi}$ over $\boldx$ one simply varies the parameters that are specific to the model (e.g. parameters in the inflationary potential --- see appendix~\ref{sec:models}) over their priors and computes the distribution over the $\boldx$ domain that this generates.

Distributions denoted with a bar --- such as $\barpi$, $\barp$ and $\barlike$ --- are defined over each individual model observable value $\boldx$, with measure $\barpi (\boldx \vert {\cal M}_\alpha ) \, \dd \boldx$ and are typically twice integrated in order to compute the expected utility: once over the $\boldx$ space and the second time over the space of $\bmuf$ so as to take into account the uncertainty in the values that a future measurement may be centred on.

\subsection*{The posterior given the future data}
Finally, we shall also consider the likelihood (defined with $\bmuf$ and $\boldsigma$) and posterior probability from a future survey, with measure $\hatp  \left[ \, \boldy \, \vert \, {\cal D}_{\rm fut}(\bmuf ,\boldsigma )  \, \right] \dd \boldy$, which is specified over the $\boldy$ (another real $n$-dimensional parameter vector sharing the same space of observables represented by $\bmuf$) domain. The futuristic dataset ${\cal D}_{\rm fut}={\cal D}_{\rm fut}(\bmuf , \boldsigma )$ is centred on $\bmuf$ with a vector of mutually independent forecast widths $\boldsigma$ which we can specify either `by hand' or through e.g. a Fisher forecasting method, given a specific survey.

All distributions denoted with a hat, such as $\hatpi$, $\hatp$ and $\hatlike$ are defined over $\boldy$. Through Bayes' rule, we can connect the posterior probability distribution given the current data (the same distribution as the one defined over $\bmuf$) to the probability distribution over the future data, once a future likelihood function has been specified
\begin{equation} \label{eq:bayes-rule-future}
\hatp  \left[ \, \boldy \, \vert \, {\cal D}_{\rm fut}(\bmuf ,\boldsigma )  \, \right] \propto  p \, (  \boldy \vert {\cal D}_{\rm cur} ) \, \hatlike \, [\, {\cal D}_{\rm fut}(\bmuf , \boldsigma ) \, \vert \, \boldy \, ] \,.
\end{equation}
Note that this distribution, and hence the points $\boldy$, are independent of the models $\boldsymbol{\calM}$. Hence, this will be useful for defining model-independent utilities later e.g. the forecast information gain. In this paper, we shall assume
\begin{equation} \label{eq:future-gaussian-assumption}
\hatlike \, [\, {\cal D}_{\rm fut}(\bmuf , \boldsigma )  \, \vert \, \boldy \, ] = {\cal N} (\boldy \vert \bmuf , \boldsigma ) \,,
\end{equation}
where the multivariate Gaussian distribution here can be defined generally as
\begin{equation}
\ {\cal N} (\boldsymbol{a} \vert \bmuf , \boldsigma ) \equiv (2\pi )^{-\frac{n}{2}} \left( \prod^n_{i=1} \sigma^i \right)^{-1} \exp \left[ - \sum^n_{i=1}\frac{(a^i-\muf^i)^2}{2(\sigma^i)^2} \right] \,,
\end{equation}
and where, crucially, we will be ignoring possible covariances. Both this and \Eq{eq:future-gaussian-assumption} will prove to be a \emph{key} assumption of this work. It is clear that forecasting for proposed missions for which the configuration of the detectors and physics of the measurement is well-understood, realistic future likelihoods may be inferred and are probably extremely complex, rendering the Gaussian assumption possibly a poor fit (we check this assumption explicitly in appendix~\ref{sec:gaussian-assumption-limitations}).

We consider this work to be a new step in developing a set of numerical forecasting tools, in which, the natural first step is to assume a Gaussian ansatz. Furthermore, we have two main reasons to focus initially on \Eq{eq:future-gaussian-assumption}:
\begin{enumerate}
\item{Our Gaussian mock forecasts represent the simplest first approximation to the full calculation where detector noises are carefully translated into error bars over the slow-roll parameters.}
\item{The narrow-variance limit of all possible $\hatlike$ distributions is well-modeled by a Dirac delta measure in $\bmuf$-space, hence the shape of our ansatz for $\hatlike$ becomes irrelevant when this limit is met (we will show that this shape-independence appears for our more futuristic surveys in \Sec{sec:results}). This is an important feature that can also be exploited for more rapid computation (see appendix~\ref{sec:foxi-computation} for further details).}
\end{enumerate}
Hence, we shall implement \Eq{eq:future-gaussian-assumption} throughout this work. A more detailed discussion of the limitations of the Gaussian assumption is provided in appendix~\ref{sec:gaussian-assumption-limitations}.

We have now clarified the important distinctions between the probability measures used within this work, so we are ready to introduce our formalism fully.

\subsection{Defining the expected utility}
To correctly manipulate our probability spaces towards the goal of this work, it is natural to define a utility function $U$ which has a dependence on the target parameters $\boldsigma$ (e.g. parameterisations of the survey geometry, as discussed in~\Ref{Bassett:2004st}). One typically seeks to maximise the expected value of $U$ in achieving a goal e.g. optimising the expected information gain from a survey with a certain configuration. Using the posterior given the current data, we can define the expected utility $\langle U\rangle$ (which can be dependent on the set of indexed models $\boldsymbol{\calM} = \{ \calM_\alpha \}$, for example) as
\begin{equation} \label{eq:exu}
\langle U \rangle = \langle U (\boldsigma ) \rangle \equiv \int_{\bmuf \in \mathbb{R}^n} U\,[\, \boldsymbol{\calM}, {\cal D}_{\rm fut} (\bmuf , \boldsigma ) \, ] \, p \,(\bmuf |{\cal D}_{\rm cur} )\,{\rm d}\bmuf \,,
\end{equation}
and, given an appropriate $U$, its corresponding centred second-moment equivalent
\begin{equation} \label{eq:secondmomentu}
\left\langle \left( U-\langle U\rangle \right)^2\right\rangle \equiv \int_{\bmuf \in \mathbb{R}^n} \bigg \{ \,  U\,[ \, \boldsymbol{\calM}, {\cal D}_{\rm fut} (\bmuf , \boldsigma ) \, \, ] -\langle U\rangle \, \bigg \}^2 \, p \,(\bmuf |{\cal D}_{\rm cur} )\,{\rm d}\bmuf \,,
\end{equation}
where $p \,(\bmuf |{\cal D}_{\rm cur} )$ is defined as the measure of uncertainty in the value that the future measurement is centred on, $\bmuf$, which is conditioned on the current data ${\cal D}_{\rm cur}$ --- which in the present case is the \emph{Planck} data. Computing both \Eq{eq:exu} and \Eq{eq:secondmomentu} above is sufficient to answer all of the questions in this work through appropriate choice of utility $U$.

\begin{figure}[t]
\begin{center}
\includegraphics[width=0.6\textwidth]{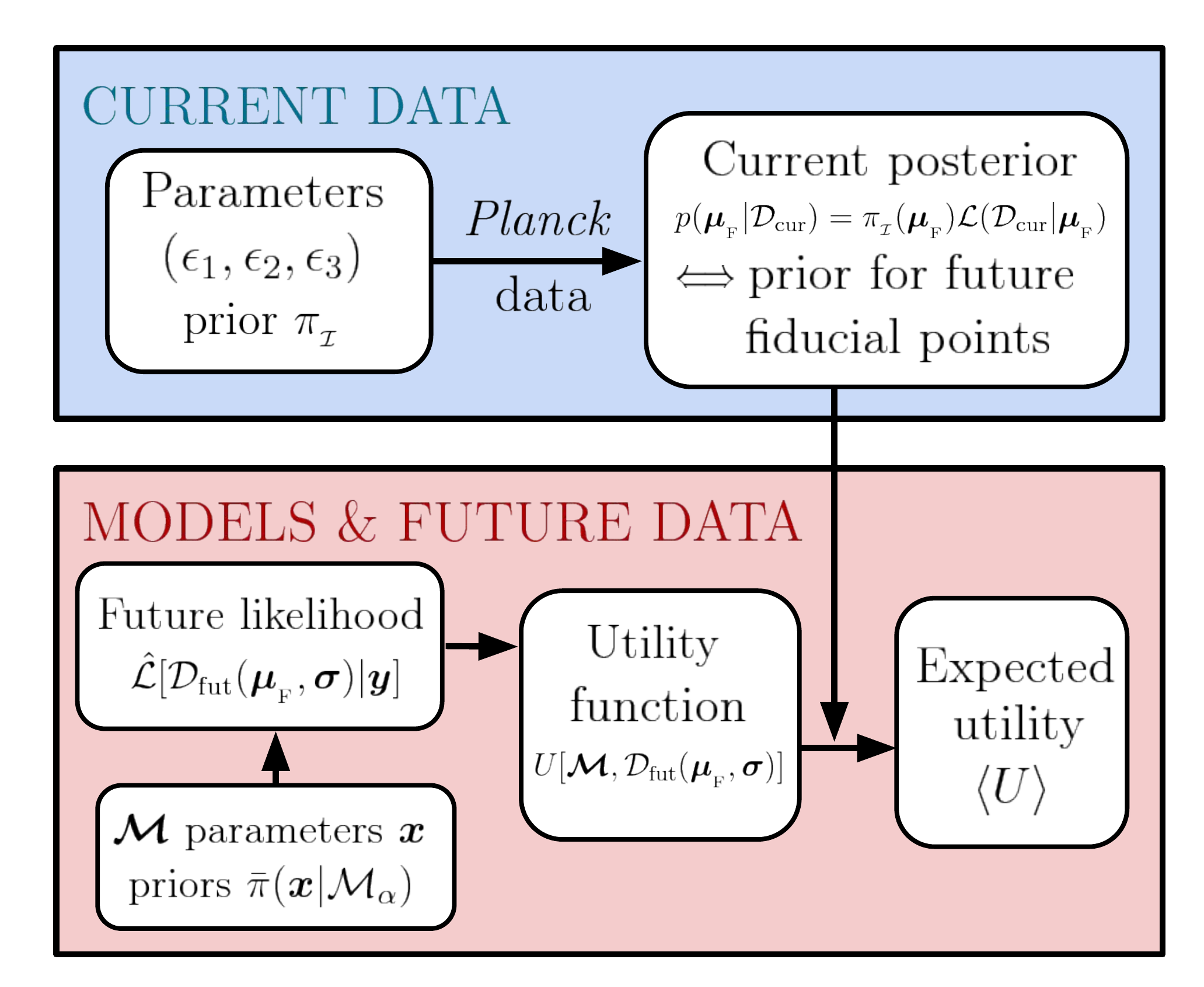}
\caption{~\label{fig:utility-diag-dec} A schematic diagram of the dependencies implied by the experimental design formalism described in \Sec{sec:formalism}. The top left hand rectangle (within the blue region) is specific to inflation --- with single-field inflationary slow roll parameters $(\epsilon_1,\epsilon_2,\epsilon_3)$ and \emph{Planck} data --- but may be replaced by any current measurement for a given survey design problem.  }
\end{center}
\end{figure}

To clarify the formalism, we have illustrated the procedure defined in this section with \Fig{fig:utility-diag-dec}. We note that the top left hand rectangle (inside the blue region), which represents the input from the {\rm Planck} data~\cite{Ade:2013sjv,Adam:2015rua,Ade:2015lrj}, may in principle be replaced with data from any measurement design problem.

\subsection{The utility functions} \label{sec:utility}
We begin by defining ${\cal E}_{\beta}$ and ${\cal E}_{\gamma}$ which denote the Bayesian evidences for two models ${\cal M}_{\beta}$ and ${\cal M}_{\gamma}$ respectively, given a future survey (and a fiducial cosmology such as $\Lambda$CDM), whose form for $\alpha = \{ \beta , \gamma \}$ is
\begin{align} \label{eq:evidences}
\ {\cal E}_\alpha (\bmuf , \boldsigma ) & \equiv \int_{\boldx \in \mathbb{R}^n} \hatlike \, [\, {\cal D}_{\rm fut}(\bmuf , \boldsigma )  \, | \, \boldx \, ] \, \barpi \,(\boldx | \calM_\alpha ) \, {\rm d} \boldx \,,
\end{align}
which uses the likelihood function $\hatlike$ from some future dataset ${\cal D}_{\rm fut}$ (assumed to be \Eq{eq:future-gaussian-assumption} in this work) defined over the model point space $\boldx$, centred at $\bmuf$ and multiplied by the prior probability measure $\barpi$ for each model.

An effective tool for model comparison is the celebrated Bayes factor ${\rm B}_{\beta \gamma}$ between two models, defined as the ratio of their evidences
\begin{equation} \label{eq:BayesFactor}
\ {\rm B}_{\beta \gamma} (\bmuf , \boldsigma )  = \frac{\cal E_{\beta} (\bmuf , \boldsigma )}{\cal E_{\gamma} (\bmuf , \boldsigma ) }\,,
\end{equation}
which favours models that realise a good compromise between quality of fit and a lack of fine tuning.\footnote{In this context, the degree of `fine-tuning' corresponds to the degree to which only a narrow region of a given models' possible observable characteristics actually fit the data well.} Thus, one favours $\calM_\beta$ within the set $\boldsymbol{\calM}$ that maximizes ${\rm B}_{\beta \gamma}$ with respect to the others. One threshold to rule ${\cal M}_{\beta}$ out with respect to ${\cal M}_{\gamma}$ is the Jeffreys threshold~\cite{jeffprob,2008arXiv0804.3173R}, where one needs to satisfy ${\cal E}_{\beta}<{\rm e}^{-5}{\cal E}_{\gamma}$. Therefore, in logarithmic terms $\ln {\rm B}_{\beta \gamma} = -5$ marks the point at which $\cal M_{\beta}$ may be considered `strongly disfavoured' versus $\cal M_{\gamma}$.

Consider now the choices of utility
\begin{align} \label{eq:BayesU}
\ U &= \left\vert \ln {\rm B}_{\beta \gamma} \right\vert \,, \\
\ U &= \Theta \left( \left\vert \ln {\rm B}_{\beta \gamma}  \right\vert -5\right) \,,\label{eq:DeciU}
\end{align}
which --- though utilities in \Eq{eq:exu} may be defined generally over the indexed model space $\boldsymbol{\calM} = \left\{ \calM_\alpha\right\}$ --- we have defined individually for each pair of models $\calM_\beta$ and $\calM_\gamma$. Depending on how observationally separable the two models are, computing the expectation value through \Eq{eq:exu} of \Eq{eq:BayesU} may provide a strong indication of the most probable absolute value of the Bayes factor, where the typical spread away from this mean value can be estimated through the centred second-moment in \Eq{eq:secondmomentu}.

Turning our attention to the other utility defined by \Eq{eq:DeciU}, the decisiveness $\deci_{\beta \gamma}$ between $\calM_\beta$ and $\calM_\gamma$, is defined as
\begin{equation}
\deci_{\beta \gamma} \equiv \left\langle \Theta \left( \vert \ln {\rm B}_{\beta \gamma}\vert-5\right) \right\rangle \,, \label{eq:decisiveness}
\end{equation}
and $\deci_{\beta \gamma}=\deci_{\gamma \beta}$, where we note that this quantity has been previously defined in \Ref{2011MNRAS.414.2337T}. $\deci_{\beta \gamma}$ incorporates the Jeffreys threshold into the decision between models, where its value is that of a real number selected from the closed interval $[0,1]$ (or the odds of a clear decision). In this way, model pairings with a large decisiveness value will be imminently distinguishable in the future, with the opposite holding true for a low decisiveness value.

Our last, model-independent,\footnote{At least dependent only upon the background cosmology.} utility function is the information gained (in the same space of observables as $\bmuf$ and $\boldx$, e.g. $(\epsilon_1,\epsilon_2,\epsilon_3)$ for our single-field inflation problem) by improving the measurement with widths $\boldsigma$ at each possible $\bmuf$
\begin{equation} \label{eq:DKLU}
\ U = \dkl \left\{ \, \hatp \left[ \, \boldy \, \vert \, {\cal D}_{\rm fut}(\bmuf ,\boldsigma )  \, \right] \, \right\vert \! \left\vert \, p \, (  \boldy \vert {\cal D}_{\rm cur} ) \, \right\} \,,
\end{equation}
also referred to as the Kullback-Leibler divergence~\cite{kullback1951} between the two distributions, which we define here as
\begin{equation}  \label{eq:calcDKL_general}
 \dkl \left\{ \, \hatp \left[ \, \boldy \, \vert \, {\cal D}_{\rm fut}(\bmuf ,\boldsigma )  \, \right] \, \right\vert \! \left\vert \, p \, (  \boldy \vert {\cal D}_{\rm cur} ) \, \right\} = \int_{\boldy \in \mathbb{R}^n} \hatp \left[ \, \boldy \, \vert \, {\cal D}_{\rm fut}(\bmuf ,\boldsigma )  \, \right] \ln \left\{ \frac{\hatp \left[ \, \boldy \, \vert \, {\cal D}_{\rm fut}(\bmuf ,\boldsigma )  \, \right] }{p \, (  \boldy \vert {\cal D}_{\rm cur} )  }\right\} \dd \boldy\,.
\end{equation}
By defining the normalisation
\begin{equation}
\ E \equiv \int_{\boldy \in \mathbb{R}^n} p \, (  \boldy \vert {\cal D}_{\rm cur} ) \, \hatlike \, [\, {\cal D}_{\rm fut}(\bmuf , \boldsigma )  \, \vert \, \boldy \, ] \, \dd \boldy \,,
\end{equation}
we can rewrite \Eq{eq:calcDKL_general}, using \Eq{eq:bayes-rule-future} and $E$, as
\begin{align}
\ & \dkl \left\{ \, \hatp \left[ \, \boldy \, \vert \, {\cal D}_{\rm fut}(\bmuf ,\boldsigma )  \, \right] \, \right\vert \! \left\vert \, p \, (  \boldy \vert {\cal D}_{\rm cur} ) \, \right\} =\nonumber \\
 & \qquad \qquad \qquad \frac{1}{E} \int_{\boldy \in \mathbb{R}^n} p \, (  \boldy \vert {\cal D}_{\rm cur} )  \, \hatlike \, [\, {\cal D}_{\rm fut}(\bmuf , \boldsigma )  \, \vert \, \boldy \, ] \ln \left\{ \frac{ \hatlike \, [\, {\cal D}_{\rm fut}(\bmuf , \boldsigma )  \, \vert \, \boldy \, ] }{E}\right\} \dd \boldy \label{eq:calcDKL}\,.
\end{align}

\subsection{The maximum-likelihood average} \label{sec:utility-expectation-values}
Throughout this work, we will use the notation $\langle \cdot \rangle$ to denote the posterior averaging as in \Eq{eq:exu}. While this is perfectly adequate to obtain expected utilities, in the case of both model-dependent utility functions (defined by \Eq{eq:BayesU} and \Eq{eq:DeciU}), one should also consider averaging over only those $\bmuf$ points that generate future likelihood distributions which do not immediately rule both models out. Indeed, in cases where both models are ruled out, the fact that one model is even more ruled out than the other does not provide valuable information and one may wish to simply discard such situations from forecasts.

An averaging scheme that can solve this problem removes the $\bmuf$ points for which the maximum likelihood of both models is too low in comparison to the global maximum likelihood. We will refer to this method hereafter as the `maximum-likelihood averaging' scheme, defined as
\begin{align}
\ &\langle \cdot \rangle_{{}_{\rm ML}} \equiv \nonumber \\
\ & \frac{1}{1-r_{{}_{\rm ML}}} \int_{\bmuf \in \mathbb{R}^n} \,\, \cdot \,\, \Theta \left[ \, \max_{i = \beta , \gamma} \left \{ \ln \hatlike \, ({\cal D}_{\rm fut}  |\boldy_* , {\cal M}_i) \right\} + t_{{}_{\rm ML}} - \ln \hatlike ({\cal D}_{\rm fut} |\bmuf ) \right]  p \, (\bmuf \vert \, {\cal D}_{\rm cur})\, \dd \bmuf \label{eq:ml-average} \,,
\end{align}
where for this work we set $t_{{}_{\rm ML}}=5$ but this threshold value can be arbitrarily defined,\footnote{Hence, we are quite restrictive, permitting only those models for which the maximum likelihood is $\hatlike \, ({\cal D}_{\rm fut}  |\boldy_* , {\cal M}_i) \geq \ee^{-5}\hatlike \, ({\cal D}_{\rm fut}  | \bmuf )$, e.g. within roughly $\sqrt{5}\simeq 2.2$-$\sigma$ of the global maximum likelihood.} we have suppressed the dependence ${\cal D}_{\rm fut} = {\cal D}_{\rm fut} (\bmuf , \boldsigma )$ for brevity and $\boldy_*$ is the maximum likelihood point for a given distribution. Thus, expected utilities generated using $\langle \cdot \rangle_{{}_{\rm ML}}$ will effectively subsample all of those possible `futures' that still require a model selection procedure to provide new information. We have also defined a normalisation factor $1-r_{{}_{\rm ML}}$ in \Eq{eq:ml-average}, where $r_{{}_{\rm ML}}$ is defined as
\begin{equation} \label{eq:rML}
\ r_{{}_{\rm ML}} \equiv \int_{\bmuf \in \mathbb{R}^n} \Theta \left[ \, \ln \hatlike ({\cal D}_{\rm fut} |\bmuf ) - t_{{}_{\rm ML}} - \max_{i = \beta , \gamma} \left \{ \ln \hatlike \, ({\cal D}_{\rm fut}  |\boldy_* , {\cal M}_i) \right\}\, \right] p \, (\bmuf \vert \, {\cal D}_{\rm cur})\, \dd \bmuf \,,
\end{equation}
hence in the limit of low accuracy $r_{{}_{\rm ML}}=0$, $\langle \cdot \rangle_{{}_{\rm ML}}=\langle \cdot \rangle$ and, in the limit of infinite accuracy, $1-r_{{}_{\rm ML}}$ measures the volume (weighted by the posterior of the current measurement) of the union of the priors between the two models. With \Eq{eq:rML} we may also keep track of the proportion of the $\bmuf$ space that has already ruled both models ${\cal M}_\beta$ and ${\cal M}_\gamma$ out with respect to the maximum likelihood point.

In \Eq{eq:decisiveness} we defined $\deci_{\beta \gamma}$ as the decisiveness between models $\calM_\beta$ and $\calM_\gamma$. Hence, using our newly developed maximum-likelihood averaging scheme in \Eq{eq:ml-average}, we define a new expected utility $\deci_{\beta \gamma}\vert_{{}_{\rm ML}}$ which we dub the `decisivity' between $\calM_\beta$ and $\calM_\gamma$. We shall make extensive use of this new quantity for the analysis \Sec{sec:results}.

\subsection{A novel computational forecasting method}
The utility functions we study here contain either of the two integrals \Eq{eq:calcDKL_general} and \Eq{eq:evidences}, which must be nested inside the integral over the $\bmuf$ point domain defined by \Eq{eq:exu} in order to compute the expected utility. The canonical approach would be to perform Nested-Nested sampling with a modification to the \texttt{MultiNest} algorithm~\cite{Feroz:2008xx}, but this would make this problem too computationally expensive due to the length of time required for (even efficient) Nested sampling to converge. Furthermore, in the particular case of the Bayes factor, we cannot always rely on the models being nested within one another, as in the implementation with the SDDR\footnote{The Savage-Dickey Density Ratio is an approximation to the Bayes factor --- valid when the models involved are nested --- which reduces the often-intractable problem of computing the Bayesian evidence to a conditional prior volume ratio.}~\cite{dickey1971, Heavens:2007ka,2011MNRAS.414.2337T}, therefore we must still perform the integrals for the evidences of each model from \Eq{eq:evidences} explicitly.

This issue can, in fact, be resolved by with a relatively simple computational programme. By relaxing the infinitessimal element in \Eq{eq:calcDKL} to be finite, we may rewrite the integral as a discrete summation
\begin{align}
 \dkl &\left\{ \, \hatp \left[ \, \boldy \, \vert \, {\cal D}_{\rm fut}(\bmuf ,\boldsigma )  \, \right] \, \right\vert \! \left\vert \, p \, (  \boldy \vert {\cal D}_{\rm cur} ) \, \right\} \simeq \nonumber \\
 & \qquad \qquad \sum_{\boldy_i \in \left\{ {\cal D}_{\rm cur} \, {\rm chains}\right\}} \hatlike_{{}_{\rm N}} \left[ \,{\cal D}_{\rm fut}(\bmuf , \boldsigma ) \, \vert \, \boldy_i \, \right] \ln \left\{ \hatlike_{{}_{\rm N}} \left[ \,{\cal D}_{\rm fut}(\bmuf , \boldsigma ) \, \vert \, \boldy_i \, \right] \right\} \label{eq:calcDKL_discrete} \,,
\end{align}
where we assume the $\boldy_i$ to be drawn from Markov chains that sample directly from $p \, (  \boldy \vert {\cal D}_{\rm cur} )$ and we have normalised the future likelihood $\hat{\like}$ in a particular way, such that
\begin{equation} \label{eq:norm_calcDKL_discrete}
\hatlike_{{}_{\rm N}} \left[ \,{\cal D}_{\rm fut}(\bmuf , \boldsigma ) \, \vert \, \boldy_i \, \right] \equiv \frac{\hatlike  \left[ \,{\cal D}_{\rm fut}(\bmuf , \boldsigma ) \, \vert \, \boldy_i \, \right]}{ \vphantom{\bigintss} \sum_{\boldy_j \in \left\{ {\cal D}_{\rm cur} \, {\rm chains}\right\}} \hatlike \left[ \, {\cal D}_{\rm fut}(\bmuf , \boldsigma ) \, \vert \, \boldy_j \, \right] }\,.
\end{equation}
Using \Eq{eq:norm_calcDKL_discrete}, \Eq{eq:calcDKL_discrete} and a sufficiently large number of points, one can efficiently compute \Eq{eq:DKLU} such that the expected utility integral in \Eq{eq:exu} --- which also must be approximated by a discrete summation --- is tractable over reasonable timescales.\footnote{$2$-$3$ days on the \href{http://www.sciama.icg.port.ac.uk/}{Sciama High Performance Compute} cluster, with $\sim 83000$ likelihood samples and $5$-$10$ models with $\sim 6000$ prior samples each.}

\Eq{eq:BayesFactor} may also be computed as a discrete summation with an appropriate weighting scheme implied by the priors of each model, where we find the following formula
\begin{equation} \label{eq:lnB-approx-discrete}
\ {\rm B}_{\beta \gamma} (\bmuf , \boldsigma ) \simeq K\frac{ \vphantom{\bigintss} \sum_{\boldx_i \in \left\{ \calM_\beta \, {\rm chains}\right\}}  \hatlike \left[ \, {\cal D}_{\rm fut}(\bmuf , \boldsigma ) \, \vert \, \boldx_i \, \right] }{ \vphantom{\bigintss} \sum_{\boldx_i \in \left\{ \calM_\gamma \, {\rm chains}\right\}}  \hatlike \left[ \, {\cal D}_{\rm fut}(\bmuf , \boldsigma ) \, \vert \, \boldx_i \, \right] } \,,
\end{equation}
in which the summations are over the Markov chains that sample directly from $\pi \, ( \boldx \vert \calM_\beta )$ (numerator) and $\pi \, ( \boldx \vert \calM_\gamma )$ (denominator) --- modulo a normalisation $K$ that exists due to varying the number of points within each chain, respectively. We note here that a  related method to compute the Bayesian evidence for the Markov chains themselves was recently introduced by \Ref{Heavens:2017hkr}, whereas the goal for this paper is forecasting with futuristic distributions which instead simplifies the integration procedure to multiple evaluations of a distribution function.

Our method can effectively construct the Bayesian evidence for any model defined by its prior over $\boldx$ and has been incorporated in our public code \href{https://github.com/umbralcalc/foxi}{\texttt{foxi}}. The algorithm to compute whichever $\langle U\rangle$ is straightforward and robust (see appendices~\ref{sec:foxi-computation} and \ref{sec:gaussian-assumption-limitations}), requiring only a minimal number of samples. The main procedure of this computation is:
\begin{enumerate}
\item{\textbf{Draw a value} from the Markov chain representing the distribution $\like ({\cal D}_{\rm cur} |\bmuf )$ and multiply its value by any necessary prior transformations to obtain $p\, (\bmuf |{\cal D}_{\rm cur})$ through \Eq{eq:post-pred}. }
\item{\textbf{Compute} the utilities $U$ using either \Eq{eq:BayesFactor} or by integrating over the whole set of posterior samples to compute the integral in \Eq{eq:calcDKL}, given the corresponding $\bmuf$ in $p\, (\bmuf |{\cal D}_{\rm cur})$. }
\item{\textbf{Store} the contribution to the integral \Eq{eq:exu}.}
\item{\textbf{If} the integral has not yet converged, go to 1.}
\item{\textbf{Compute} \Eq{eq:exu} and \Eq{eq:secondmomentu} using the contributions stored in 3.}
\end{enumerate}
In order to calculate expected utilities with the $\langle \cdot \rangle_{{}_{\rm ML}}$ average, one simply discards points at steps 1. and 4. which do not satisfy the condition within \Eq{eq:ml-average}. We also note that higher-order statistics such as \Eq{eq:secondmomentu} can be computed trivially from the samples generated by this algorithm.

We shall now progress to analyse the results obtained for the surveys introduced in Table~\ref{tab:future-widths}. We refer the interested reader to appendix~\ref{sec:foxi-computation} for further details on the computational strategies and robustness checks we have implemented in the code.
\section{Results and analysis}
\label{sec:results}
In all of the analysis below we will consider probability distributions over the various utilities defined in the previous section given a set of futuristic measurement widths. In Table \ref{tab:future-widths} we listed the different settings used for each futuristic scenario, where in each case we represented the characteristic measurement errors that might be forecast for a particular configuration of experiment. The specifications of the first two experiments are relatively close to being realised by either CMB Stage-4~\cite{Abazajian:2016yjj}, LiteBIRD~\cite{litebird} or  COrE~\cite{DiValentino:2016foa,Finelli:2016cyd} and are therefore optimistically labeled `Proposed' with P1 (CMB Stage-4) and P2 (LiteBIRD/COrE). The other four configurations represent a futuristic order of magnitude improvement in the constraint on each of the three slow roll parameters (F1-3), where the final one represents the simultaneous improvement in all three previous configurations (F4).

\begin{figure}[t]
\begin{center}
\includegraphics[width=0.47\textwidth]{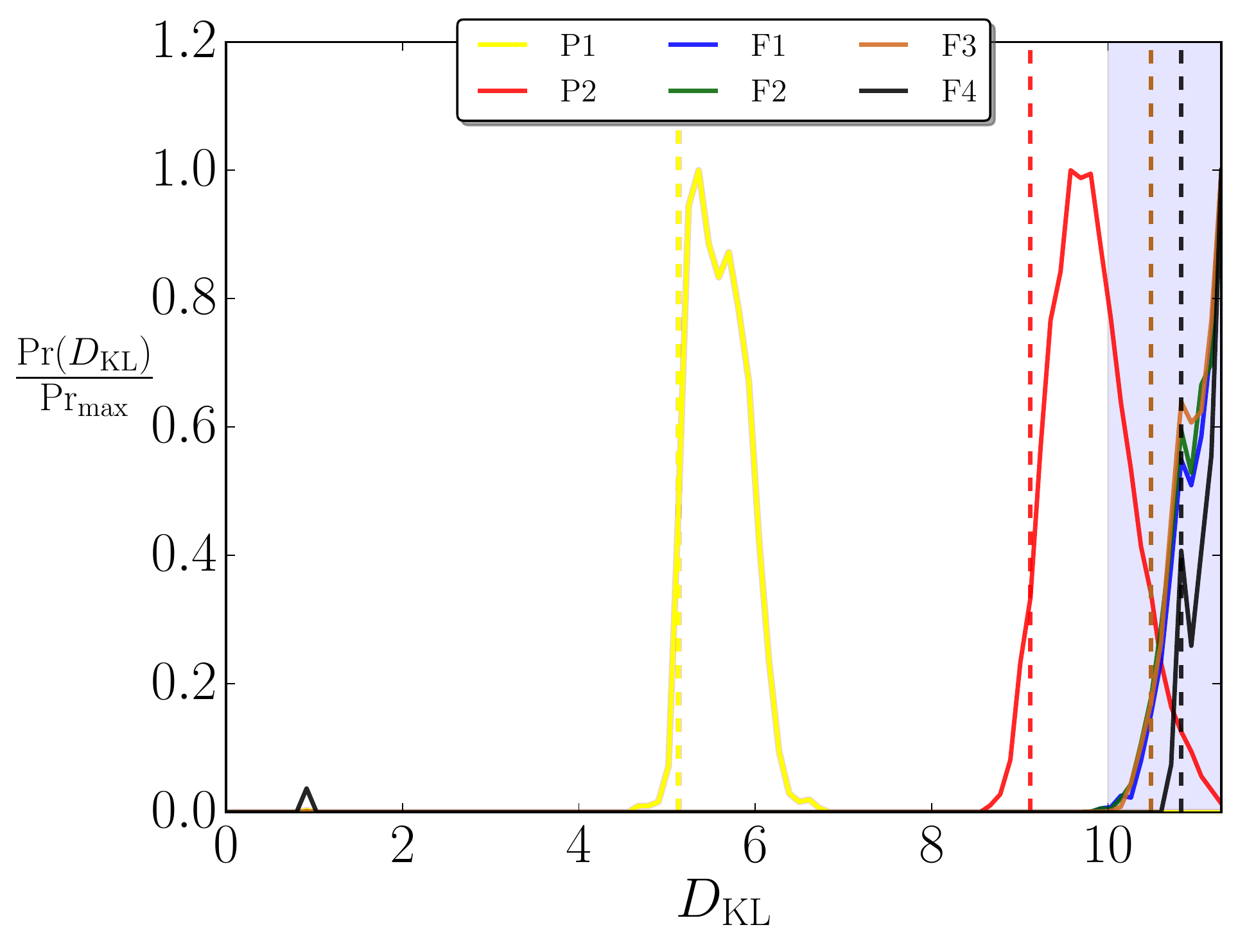}
\includegraphics[width=0.47\textwidth]{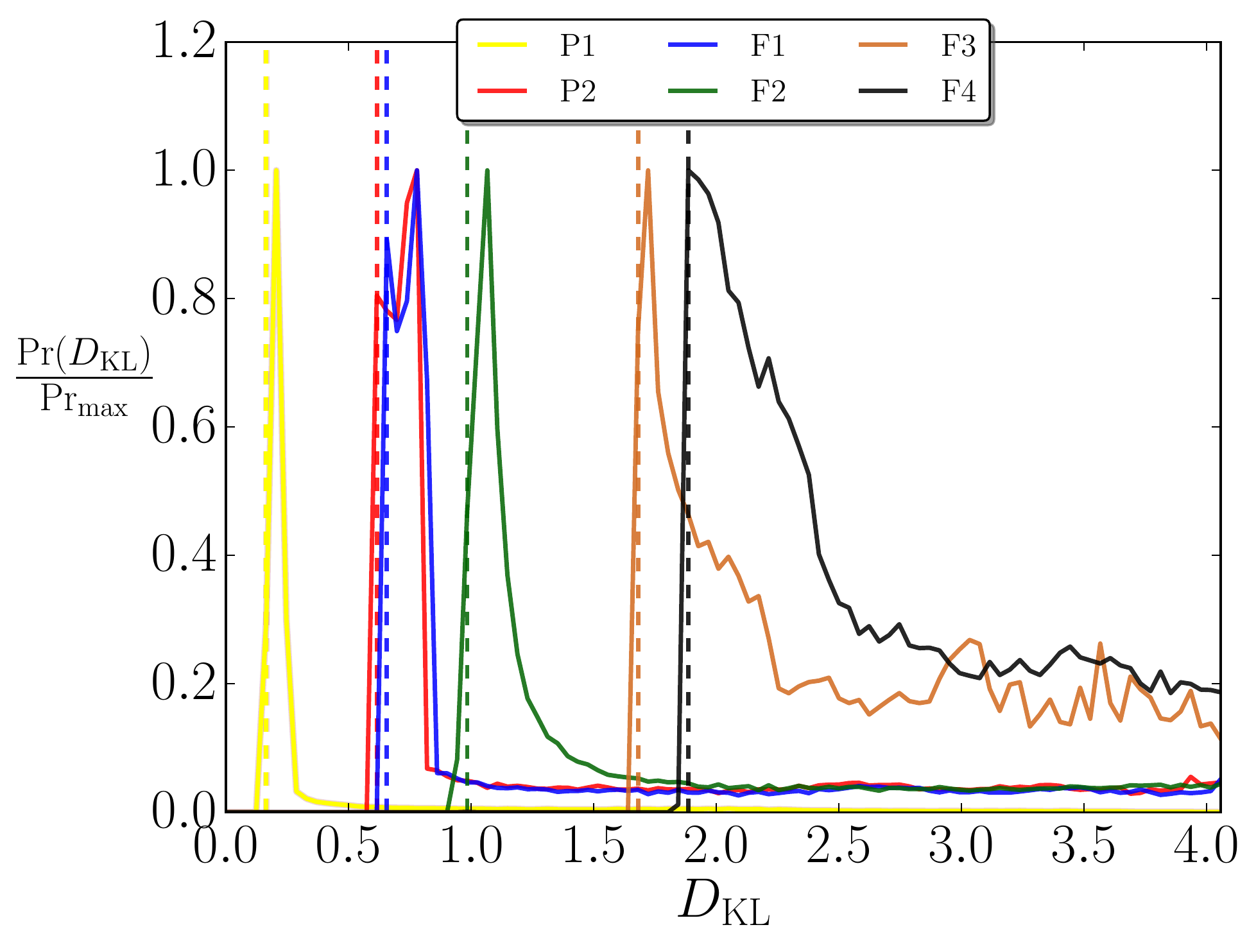}
\caption{~\label{fig:info-gain-future} Binned probability density plots showing the distribution of values of the Kullback-Leibler divergence $\dkl$ corresponding to each set of futuristic widths in Table \ref{tab:future-widths}. The vertical line associated to each colour is the 95\% lower bound for each experiment. The posterior samples are derived from the Planck data marginalised using the Machine Learning methods defined in~\Ref{Ringeval:2013lea} over $( \epsilon_1 , \epsilon_2, \epsilon_3 )$. The plot on the left uses the $\pi (\bmuf \vert \, \epsilon_1 )$ prior (see \Eq{eq:predictive-priors-flateps1}) where one can see that $\dkl$ in this case is predominately $>11.4$ for F1-4. The plot on the right assumes the $\pi (\bmuf \vert \log \epsilon_1 )$ prior (see \Eq{eq:predictive-priors-logeps1}). The grey region in the plot on the left side represents the region $\dkl > 11.4$ beyond the precision of our numerical procedure (see main text). }
\end{center}
\end{figure}

In Table \ref{tab:future-widths} we have also displayed the expectation value on the $\dkl$ (information gain) between the current Planck data and each future dataset in turn. The 95\% bound in each case is also depicted with the dashed lines in \Fig{fig:info-gain-future} where the solid lines represent the marginalised probability density in the future of the $\dkl$ value. The distinction between a choice of prior is striking (left and right plots correspond to \Eq{eq:predictive-priors-flateps1} and \Eq{eq:predictive-priors-logeps1} respectively) where e.g. all of the F1-4 datasets saturate an effective numerical upper bound on the expected information gain achievable $\langle \dkl \rangle > 11.4$. Notice indeed that \Eq{eq:norm_calcDKL_discrete} is limited by the number of samples in the Markov chains representing ${\cal D}_{\rm cur}$, such that the typical number of samples used for computations over this space in this paper ($\sim 85000$) yields this upper bound directly $\ln (85000) \simeq 11.4$.

The value of $\langle \dkl \rangle$ appears to rise far more quickly towards the numerical bound in the case of the flat prior over $\epsilon_1$ as opposed to logarithmic $\epsilon_1$, which can be attributed to the improvements in measurement errors that squeeze up to the hard prior lower bound in the former case, which is $\epsilon_1 \geq 10^{-4}$ from \Eq{eq:predictive-priors-flateps1}. Due to this strong hard prior bound dependence there is a large information gain, which is to be expected when the measurement precision over $\epsilon_1$ becomes of the same order as this bound. From \Fig{fig:nsr-plot} one can also see that two of the models are already ruled out by such a measurement (KMIII and ${\rm KKLTI}_{\rm stg}$) due to their tensor-to-scalar ratios (given by $16\epsilon_1$, see \Eq{eq:tens-scalar-ratio}) being both orders of magnitude below this bound. For this reason we will only consider the logarithmic prior over $\epsilon_1$ defined by \Eq{eq:predictive-priors-logeps1} when considering our model selection utilities, since it is a far more conservative choice.

Turning our attention now to the values of $\dkl$ sampled by the $\bmuf$ points using a logarithmic prior over $\epsilon_1$ in \Fig{fig:info-gain-future}, we see a clear trend and increase in information gain by each survey configuration, which is matched by the values of $\langle \dkl \rangle$ in Table \ref{tab:future-widths}. Notably, the optimal information gain between surveys F1-3 is achieved through improvements to the measurement over $\epsilon_3$ in F3. This is clearly due to the fact that the current constraints are the least constraining over $\epsilon_3$ when compared with the other two parameters in the slow-roll hierarchy. We shall return to this interesting point for further discussion in \Sec{sec:concl}.
\subsection{General statements}
\label{sec:gen-state}
The combined results of this paper span Tables \ref{tab:p1-utilities}, \ref{tab:p2-utilities} and \ref{tab:f14-utilities}. We have performed the analysis computing $\langle \vert \ln {\rm B}_{\beta \gamma} \vert \rangle$, $\deci_{\beta \gamma}$, $\langle \vert \ln {\rm B}_{\beta \gamma} \vert \rangle_{{}_{\rm ML}}$ and $\deci_{\beta \gamma}\vert_{{}_{\rm ML}}$ as expected utilities using all possible pairs of the models defined in appendix~\ref{sec:models}, where the latter two expected utilities make use of the maximum-likelihood average $\langle \cdot \rangle_{{}_{\rm ML}}$ from \Eq{eq:ml-average}. In addition, we have also provided the ratio of rejected points $r_{{}_{\rm ML}}$
according to this alternative averaging scheme defined by \Eq{eq:rML} in each table.

The increasing decisivity between models is best summarised in \Fig{fig:allmodels_deci}, where the general trend begins with survey P1, where no value of $\deci_{\beta \gamma}\vert_{{}_{\rm ML}}$ is above a probability of 0.1,  towards complete certainty of a decision between all model pairs ($\deci_{\beta \gamma}\vert_{{}_{\rm ML}}=1.0$) in survey F4. An important detail to note at this point is that between F1-3 the best decisive outcome between all model pairs is achieved by survey F2, which corresponds to an order of magnitude decrease in the measurement errors over the second slow-roll parameter $\epsilon_2$. This already gives a strong indication that the possible future directions for selection between inflationary models may rely more on increased precision over the spectral index $\nS$ and less on the tensor-to-scalar ratio $r$. We shall, once again, return to this discussion point later in \Sec{sec:concl}.

\subsection{Forecasts using P1 and P2}
\label{sec:p1p2}

We first examine Tables \ref{tab:p1-utilities} and \ref{tab:p2-utilities} (P1 and P2 surveys, respectively corresponding to CMB Stage-4 and COrE/LiteBIRD-like surveys) which use the measurement widths that are expected to be achievable in the relatively near future, whence, the label `P' for `Proposed'. For P1 the $r_{{}_{\rm ML}}$ values suggest that already $\sim 2-4\%$ of the possible future realisations will rule both models of each pair out at the level of either model's maximum likelihood given our threshold of $\ee^{-t_{{}_{\rm ML}}}\like_{\rm max}$ or above (see \Sec{sec:utility-expectation-values}), where $t_{{}_{\rm ML}}=5$. Note that this is not the same as \emph{all} of the model pairs being ruled out at once but instead reflects the specific decision question for each model in-turn. P2 has a far more striking result --- in $\geq 94\%$ of the possible future measurements, both models in each pair (in all 10 possible combinations) will have been eliminated at the maximum likelihood level. We can infer from these results alone that the upcoming future surveys of the P2-type will have strong decision-making capabilities even before any further analysis or detailed model selection program is initiated. This indicates that an important first threshold in the space of possible CMB missions exists, somewhere between the capabilities of P1 and P2, where most single-field model pairs will already be ruled out at the level of their maximum likelihoods. This threshold can be crossed in the future by a COrE/LiteBIRD-like mission.

Let us move on to the expected model selection utilities by improving measurement bounds by an order of magnitude on both $\epsilon_1$ and $\epsilon_3$. In doing so we advance from P1 to P2, where most model pairs receive a very large amplitude increase in $\langle \vert \ln {\rm B}_{\beta \gamma} \vert \rangle_{{}_{\rm ML}}$ e.g. all of the pairs that include the RGI model increase by an order of magnitude in $\ln$-scale. The uncertainties associated to this expected utility also become significantly larger in most cases. Though it is instructive to consider the expected Bayes factor utilities, the variance in their value for each model pair (especially in the case of survey P2) leads to significant uncertainty in assertions about the future that rely on these utilities alone. Therefore, we can support our claims by considering the decisivity $\deci_{\beta \gamma} \vert_{{}_{\rm ML}}$ for the same pairs of models, where most receive a greater-than factor of 4 increase in the odds of a decisive model selection with survey P2 when compared to P1.

\begin{table}
\centering
\begin{tabular}{|c|c|c|c|c|c|}
\cline{2-6}
\multicolumn{1}{c}{\cellcolor{yellow!55}} & \multicolumn{5}{|c|}{P1 with $\pi (\bmuf \vert \log \epsilon_1)$}  \\      \hline
\backslashbox{${\cal M}_\beta$ - ${\cal M}_\gamma$}{$\langle  U\rangle$}  & $\langle \vert \ln {\rm B}_{\beta \gamma}\vert \rangle$ & $\langle \vert \ln {\rm B}_{\beta \gamma} \vert \rangle_{{}_{\rm ML}}$ & $\mathscr{D}_{\beta \gamma}$ & $\mathscr{D}_{\beta \gamma}\vert_{{}_{\rm ML}}$ & $r_{{}_\mathrm{ML}}$ \\     \hline\hline
${\rm KMIII}$ - ${\rm HI}$ & $2.42 \,(< 91.72)$ & $2.41 \,(< 92.88)$ & 0.01 & $0.0+\varepsilon$ & 0.04  \\      \hline
${\rm KKLTI}_{\rm stg}$ - ${\rm HI}$ & $3.20 \,(< 52.88)$ & $3.22 \,(< 53.36)$ & 0.03 & 0.03 & 0.03  \\      \hline
${\rm LI}_{\alpha >0}$ - ${\rm HI}$ & $3.21 \,(< 17.24)$ & $2.99 \,(\pm 1.15)$ & 0.06 & 0.04 & 0.03  \\      \hline
${\rm RGI}$ - ${\rm HI}$ & $3.09 \,(< 61.96)$ & $1.41 \,(< 4.74)$ & 0.01 & 0.01 & 0.03  \\      \hline
${\rm KKLTI}_{\rm stg}$ - ${\rm KMIII}$ & $5.33 \,(< 104.30)$ & $5.42 \,(< 105.64)$ & 0.03 & 0.03 & 0.03  \\      \hline
${\rm LI}_{\alpha >0}$ - ${\rm KMIII}$ & $5.59 \,(< 93.06)$ & $5.39 \,(< 92.26)$ & 0.08 & 0.06 & 0.03  \\      \hline
${\rm RGI}$ - ${\rm KMIII}$ & $5.48 \,(< 110.64)$ & $3.79 \,(< 92.22)$ & 0.03 & 0.01 & 0.02  \\      \hline
${\rm LI}_{\alpha >0}$ - ${\rm KKLTI}_{\rm stg}$ & $5.03 \,(< 55.04)$ & $4.85 \,(< 52.82)$ & 0.07 & 0.04 & 0.04  \\      \hline
${\rm RGI}$ - ${\rm KKLTI}_{\rm stg}$ & $5.04 \,(< 81.50)$ & $3.40 \,(< 53.26)$ & 0.04 & 0.03 & 0.03  \\      \hline
${\rm RGI}$ - ${\rm LI}_{\alpha >0}$ & $3.46 \,(< 59.12)$ & $1.83 \,(< 4.34)$ & 0.03 & 0.01 & 0.04  \\      \hline
\end{tabular}
\caption{~\label{tab:p1-utilities} Computed expected utilities for a P1 experiment. All results correspond to a choice of the $\pi (\bmuf \vert \log \epsilon_1 )$ prior in \Eq{eq:predictive-priors-logeps1}. Note that $\varepsilon$ reminds the reader that the value is subject to rounding errors of up to $0.005$. Values in brackets $\pm$ around each computed expected utility correspond to the $1$-$\sigma$ errors, which are evaluated using \Eq{eq:secondmomentu}. This symmetric error about our different expected values for $\vert \ln {\rm B}_{\beta \gamma} \vert$ is replaced with a $2$-$\sigma$ upper bound (because it is positive by definition) if the lower error is greater than the expected value itself.}
\end{table}

\begin{table}
\centering
\begin{tabular}{|c|c|c|c|c|c|}
\cline{2-6}
\multicolumn{1}{c}{\cellcolor{red!55}} & \multicolumn{5}{|c|}{P2 with $\pi (\bmuf \vert \log \epsilon_1 )$}  \\      \hline
\backslashbox{$\calM_\beta$ - $\calM_\gamma$}{$\langle  U\rangle$}  & $\langle \vert \ln {\rm B}_{\beta \gamma}\vert \rangle$ & $\langle \vert \ln {\rm B}_{\beta \gamma} \vert \rangle_{{}_{\rm ML}}$ & $\deci_{\beta \gamma}$ & $\deci_{\beta \gamma}\vert_{{}_{\rm ML}}$ & $r_{{}_\mathrm{ML}}$ \\     \hline\hline
${\rm KMIII}$ - ${\rm HI}$ & $10.04 \,(< 105.64)$ & $43.76 \,(< 391.02)$ & 0.79 & 0.12 & 0.95  \\      \hline
${\rm KKLTI}_{\rm stg}$ - ${\rm HI}$ & $10.16 \,(< 55.06)$ & $3.06 \,(\pm 2.23)$ & 0.87 & 0.11 & 0.94  \\      \hline
${\rm LI}_{\alpha >0}$ - ${\rm HI}$ & $6.10 \,(< 77.44)$ & $2.42 \,(< 6.80)$ & 0.09 & 0.09 & 0.96  \\      \hline
${\rm RGI}$ - ${\rm HI}$ & $15.49 \,(< 185.90)$ & $18.73 \,(< 222.02)$ & 0.69 & 0.06 & 0.95  \\      \hline
${\rm KKLTI}_{\rm stg}$ - ${\rm KMIII}$ & $4.77 \,(< 91.26)$ & $39.57 \,(< 378.58)$ & 0.08 & 0.16 & 0.94  \\      \hline
${\rm LI}_{\alpha >0}$ - ${\rm KMIII}$ & $9.98 \,(< 133.10)$ & $41.38 \,(< 378.46)$ & 0.12 & 0.22 & 0.94  \\      \hline
${\rm RGI}$ - ${\rm KMIII}$ & $15.51 \,(< 214.72)$ & $55.34 \,(< 430.48)$ & 0.05 & 0.11 & 0.94  \\      \hline
${\rm LI}_{\alpha >0}$ - ${\rm KKLTI}_{\rm stg}$ & $9.98 \,(< 97.56)$ & $3.34 \,(< 12.84)$ & 0.65 & 0.11 & 0.94  \\      \hline
${\rm RGI}$ - ${\rm KKLTI}_{\rm stg}$ & $15.62 \,(< 194.58)$ & $19.28 \,(< 224.84)$ & 0.11 & 0.13 & 0.94  \\      \hline
${\rm RGI}$ - ${\rm LI}_{\alpha >0}$ & $10.68 \,(< 169.88)$ & $13.89 \,(< 198.60)$ & 0.04 & 0.04 & 0.95  \\      \hline
\end{tabular}
\caption{~\label{tab:p2-utilities} Computed expected utilities for a P2 experiment. All results correspond to a choice of the $\pi (\bmuf \vert \log \epsilon_1 )$ prior in \Eq{eq:predictive-priors-logeps1}. Note that $\varepsilon$ reminds the reader that the value is subject to rounding errors of up to $0.005$. Values in brackets $\pm$ around each computed expected utility correspond to the $1$-$\sigma$ errors, which are evaluated using \Eq{eq:secondmomentu}. This symmetric error about our different expected values for $\vert \ln {\rm B}_{\beta \gamma} \vert$ is replaced with a $2$-$\sigma$ upper bound (because it is positive by definition) if the lower error is greater than the expected value itself.}
\end{table}

\subsection{Forecasts using F1-4}
\label{sec:f14}

We begin our analysis of the results using surveys F1-4 in Table \ref{tab:f14-utilities} by noting that, from this point onward, because the measurement errors for each survey are so small it will no longer be informative to use $\langle \vert \ln {\rm B}_{\beta \gamma}\vert \rangle$ and  $\langle \vert \ln {\rm B}_{\beta \gamma}\vert \rangle_{{}_{\rm ML}}$ since their magnitudes are all above the Jeffrey's threshold $>5$ (and probably above the numerical precision). It is, however, far more illuminating to examine the values of $\deci_{\beta \gamma} \vert_{{}_{\rm ML}}$ and $r_{{}_{\rm ML}}$ together: firstly to assert whether or not the proportion of $\bmuf$ points remaining is already very small for which Bayesian model selection techniques are unnecessary (i.e. how large $r_{{}_{\rm ML}}$ is will dictate how likely it is in the future for a given model pair to be totally ruled out at the level of the maximum likelihood, and hence whether there are any likely futures for which Bayesian model selection will be required at all), and secondly in the event of model selection being required, whether or not $\deci_{\beta \gamma} \vert_{{}_{\rm ML}}$ gives good odds of successfully deciding between those models.

Survey F1 increases the measurement precision over $\epsilon_1$ from P2 by an order of magnitude. Using Table \ref{tab:f14-utilities}, for each pair of models this improvement is expected to leave a $\leq 0.06$ chance of avoiding a ruling-out with respect to the maximum likelihood of each model. Of the expected remaining $\bmuf$ points, there is varied performance by Bayesian model selection to be decisive --- one the one hand, KMIII - HI and ${\rm KKLTI}_{\rm stg}$ - HI are always decided between ($\deci_{\beta \gamma} \vert_{{}_{\rm ML}} = 1.0 - \varepsilon$ up to rounding errors $\varepsilon = 0.005$), whereas on the other hand, there are only chances of 0.12 and 0.18 to decide between RGI - ${\rm LI}_{\alpha >0}$ and RGI - ${\rm KKLTI}_{\rm stg}$, respectively.

In contrast, survey F2 increases the measurement precision over $\epsilon_2$ from P2 by an order of magnitude. For this improvement, one lowers slightly further the chance of avoiding a ruling-out with respect to the maximum likelihood of each model down to $\leq 0.05$.  Of the expected remaining $\bmuf$ points, there is a very impressive performance expected, yielding at worst chances of 0.47 and 0.5 to decide between the  pairs KMIII - HI and RGI - HI (also ${\rm KKLTI}_{\rm stg}$ - HI) respectively where, in fact, most other model pairs have high decisivity $\geq 0.68$. It is for this reason that we will conclude later that an F2 strategy for survey design is superior to F1 for single-field inflationary model selection.

Survey F3 increases the measurement precision over $\epsilon_3$ from P2 by an order of magnitude. Between F1-3 this survey configuration has the greatest chance of ruling out a given model pair at the level of the maximum likelihood, which is $\geq 0.96$. Of the remaining $\bmuf$ points, there is a wildly varied chance of a decisive conclusion between models e.g. 0.12 for RGI - ${\rm LI}_{\alpha >0}$, but conversely, a chance of $\geq 0.76$ for all model pairs including ${\rm KKLTI}_{\rm stg}$.

The decisiveness $\deci_{\beta \gamma}$ drops dramatically from F1 and F2 to F3 (and also F4 which inherits this feature from F3). This is as a feature that arises from situations where the Bayesian evidence of both models being too low to numerically evaluate, and hence the algorithm assigns $\vert \ln {\rm B}_{\beta \gamma} \vert = 0$, which results in a contribution of 0 to the decisiveness at that point. If this happens frequently enough then the value of $\deci_{\beta \gamma}$ drops accordingly, as is the case when the measurement precision over $\epsilon_3$ is improved enough for it to be a decisive observable. In principle this can be rectified by hand by assuming that $\vert \ln {\rm B}_{\beta \gamma} \vert > 5$ for all of these points, but this is not strictly correct, and hence we have not quoted $\deci_{\beta \gamma}$ for F3 and F4 accordingly. This numerical problem does not exist for the decisivity $\deci_{\beta \gamma} \vert_{{}_{\rm ML}}$, and hence provides another supporting argument for its use.

Finally, because using F4 always appears to give values of $r_{{}_{\rm ML}}\geq 0.97$, we can immediately conclude that the survey configuration F4 is close to the ultimate goal for, essentially, absolute certainty in deciding between the plateau models at the level of their maximum likelihood values alone. The fact that $r_{{}_{\rm ML}}$ saturates to a constant value for most model pairs in moving from F1-3 to F4 indicates that there is a second threshold in the space of CMB missions (the first being between P1 and P2). The value of $r_{{}_{\rm ML}}$ saturates to a constant when the measurement over $(\epsilon_1,\epsilon_2,\epsilon_3)$ is so precise that it is effectively a Dirac delta function when compared with the priors over a pair of models. Hence, the value of $1-r_{{}_{\rm ML}}$ in this limit (as discussed previously in \Sec{sec:utility-expectation-values}) corresponds to the total prior union volume of the two models relative to the total volume in the $(\epsilon_1,\epsilon_2,\epsilon_3)$ space that is weighted by the current likelihood $\like ({\cal D}_{\rm cur}\vert \bmuf )$.

Furthermore, in this limit, the Bayes factor between all model pairs reduces to a trivial prior point ratio
\begin{equation} \label{eq:trivial-bayes-factor}
\left. {\rm B}_{\beta \gamma} \right\vert_{\boldsigma \rightarrow 0} \rightarrow \frac{ \vphantom{\bigintss} \bigintsss_{\boldx \in \mathbb{R}^n} \delta (\boldx - \bmuf) \, \barpi \, ( \boldx | \calM_\beta )\, \dd \boldx  }{ \vphantom{\bigintss} \bigintsss_{\boldx \in \mathbb{R}^n} \delta (\boldx - \bmuf ) \, \barpi \, ( \boldx | \calM_\gamma )\, \dd \boldx  } = \frac{\barpi \, ( \bmuf | \calM_\beta )  }{ \barpi \, ( \bmuf | \calM_\gamma ) } \,,
\end{equation}
and note that this becomes independent of the future measurement widths $\boldsigma$. Hence, to go any further than this measurement precision will require a reformulation of a new space of models $\boldsymbol{\calM}$ with priors that are coarse-grained to much finer detail so as to remain competitive.

\subsection{Deciding between reheating scenarios}
\label{sec:reheating}

Full statistical inference of the temperature of reheating for a given inflationary model is an exciting new research topic within early Universe cosmology~\cite{Martin:2014nya,Hardwick:2016whe,Finelli:2016cyd,Martin:2016oyk}. In principle, if one can infer a micro-physical parameter, such as temperature, from the thermal bath at high energies then the early Universe can become a laboratory for high-energy physics. In addition to this, one can potentially distinguish between inflationary models with the same potential, e.g. Higgs inflation~\cite{Bezrukov:2007ep} and Starobinsky inflation~\cite{Starobinsky:1992ts}, that are realised in different theoretical frameworks by using their possibly different reheating temperatures.

In this short section we use our formalism to study 3 nested models within the HI model: ${\rm HI}_{T-}$, ${\rm HI}_{T}$ and ${\rm HI}_{T+}$, which correspond to the HI potential at fixed reheating temperatures $T_{\rm reh}=10^{12}\, {\rm GeV}$, $10^{6}\, {\rm GeV}$ and $1\, {\rm GeV}$, respectively. Motivations for the reheating temperatures include the various relic species overproduction problems, e.g., the so-called `gravitino problem'~\cite{Dimastrogiovanni:2015wvk} for the lower temperature at $T_{\rm reh}= 1{\rm GeV}$, reheating temperatures of $T_{\rm reh}= 10^6{\rm GeV}$ are favoured by Supergravity channels for Starobinsky inflation~\cite{Terada:2014uia} and $T_{\rm reh}=10^{12}{\rm GeV}$ is typical for Higgs inflation~\cite{2009JCAP06029B}.

By performing the same analysis to compute the expected utilities for the comparison between these nested models, we will give a qualitative impression of how our formalism can be used to indicate the future performance of any survey with respect to carrying out inference on reheating.

Table \ref{tab:reheating-temps} lists our full results for this analysis. The chance of ruling out all of the reheating temperatures at the level of the maximum likelihood reaches 1.0 with surveys F1-4, and the reheating temperatures are essentially measured to extremely good precision, therefore we have not included these results in the table since they are essentially trivial.

Considering the results using the P1 configuration first, the chance of ruling out each pair of temperatures at the level of the maximum likelihood is low ($\leq 0.05$). In addition, we find that model selection offers no additional benefit of deciding between temperatures for the HI model since $\langle \vert \ln {\rm B}_{\beta \gamma} \vert \rangle_{{}_{\rm ML}}$ is well below $5$ (even with the typical standard deviation added) and $\deci_{\beta \gamma}\vert_{{}_{\rm ML}}$ supports this by indicating a 0.0 (up to rounding errors of 0.005) chance of decisive selection of temperature.

We now turn our attention to the P2 configuration. According to Table \ref{tab:reheating-temps}, the improvements to the measurement bounds in moving from P1 to P2 indicate that one can nearly be certain (chance of $\geq 0.97$) that they will be able to select away from each pair of reheating temperatures at the level of the maximum likelihood, boding well in this regard for the prospects of future surveys like COrE\footnote{In addition, supporting the conclusions made by \Ref{Finelli:2016cyd}}~\cite{Finelli:2016cyd}.

If one now considers the values of the $\langle \vert \ln {\rm B}_{\beta \gamma}\vert \rangle_{{}_{\rm ML}}$ utility for the P2 survey, these suggest that future values of $\vert\ln {\rm B}_{\beta \gamma}\vert \simeq 2$ occur more regularly at $2$-$\sigma$ for all three reheating temperatures, and hence they may be distinguished between, which is indeed consistent with \Ref{Finelli:2016cyd}. We note, however that this does not mean that such temperatures can be decisively ruled out with respect to one another --- a fully decisive future with $\vert\ln {\rm B}_{\beta \gamma}\vert = 5$ appears to occur only very infrequently at the beyond 5-$\sigma$ level.

We have demonstrated the versatility that our formalism has, as well as the range of applicable problems that the \href{https://sites.google.com/view/foxicode}{\texttt{foxi}} package can deal with. We continue to the next section with another example.

\begin{table}
\centering
\begin{tabular}{|c|c|c|c|c|c|c|c|}
\cline{1-8}
\multicolumn{2}{|c|}{Survey} &\backslashbox{$\calM_\beta$ - $\calM_\gamma$}{$\langle  U\rangle$}  & $\langle \vert \ln {\rm B}_{\beta \gamma}\vert \rangle$ & $\langle \vert \ln {\rm B}_{\beta \gamma} \vert \rangle_{{}_{\rm ML}}$ & $\deci_{\beta \gamma}$ & $\deci_{\beta \gamma}\vert_{{}_{\rm ML}}$ & $r_{{}_\mathrm{ML}}$ \\     \hline\hline
\cellcolor{yellow!55} & P1  & ${\rm HI}_{T-}$ - ${\rm HI}_{T}$ & $0.39 \,(\pm 0.30)$ & $0.35 \,(\pm 0.20)$ & $0.0+\varepsilon$ & $0.0+\varepsilon$ & 0.05  \\      \hline
\cellcolor{yellow!55} & P1  & ${\rm HI}_{T-}$ - ${\rm HI}_{T+}$ & $0.79 \,(\pm 0.55)$ & $0.72 \,(\pm 0.38)$ & $0.0+\varepsilon$ & $0.0+\varepsilon$ & 0.04  \\      \hline
\cellcolor{yellow!55} & P1  & ${\rm HI}_{T}$ - ${\rm HI}_{T+}$ & $0.41 \,(\pm 0.25)$ & $0.37 \,(\pm 0.17)$ & $0.0+\varepsilon$ & $0.0+\varepsilon$ & 0.05  \\      \hline
\cellcolor{red!55} & P2 & ${\rm HI}_{T-}$ - ${\rm HI}_{T}$ & $2.09 \,(< 17.28)$ & $1.61 \,(\pm 0.52)$ & 0.04 & $0.0+\varepsilon$ & 0.98  \\      \hline
\cellcolor{red!55} & P2 & ${\rm HI}_{T-}$ - ${\rm HI}_{T+}$ & $4.17 \,(< 29.86)$ & $2.72 \,(\pm 0.93)$ & 0.12 & $0.0+\varepsilon$ & 0.97  \\      \hline
\cellcolor{red!55} & P2 & ${\rm HI}_{T}$ - ${\rm HI}_{T+}$ & $2.09 \,(< 24.12)$ & $1.0 \,(\pm 0.39)$ & 0.02 & $0.0+\varepsilon$ & 0.97  \\      \hline
\end{tabular}
\caption{\label{tab:reheating-temps} Computed expected utilities for the Higgs Inflation (HI) model (defined by the potential of \Eq{eq:higgs-pot}) fixed with 3 different reheating temperatures, where ${\rm HI}_{T-}$, ${\rm HI}_{T}$ and ${\rm HI}_{T+}$ each correspond to the model with reheating temperatures $T_{\rm reh}=1\, {\rm GeV}$, $10^{6}\, {\rm GeV}$ and $10^{12}\, {\rm GeV}$, respectively. The expected utilities have been computed with the first 2 survey configurations studied in this paper (P1 and P2) and all results correspond to a choice of the $\pi (\bmuf \vert \log \epsilon_1 )$ prior in \Eq{eq:predictive-priors-logeps1}. Note that $\varepsilon$ reminds the reader that the value is subject to rounding errors of up to $0.005$. Values in brackets $\pm$ around each computed expected utility correspond to the $1$-$\sigma$ errors, which are evaluated using \Eq{eq:secondmomentu}. This symmetric error about our different expected values for $\vert \ln {\rm B}_{\beta \gamma} \vert$ is replaced with a $2$-$\sigma$ upper bound (because it is positive by definition) if the lower error is greater than the expected value itself. }
\end{table}

\subsection{Measuring the scalar running}
\label{sec:measuring-alphaS}

Another example of our formalism at work is in the forecasting of the probability that as-of-yet unobserved parameters will be measured in the future by a given survey with forecast widths $\boldsigma$. Consider the running\footnote{This is also a good consistency check with our assumption that the $(\epsilon_1,\epsilon_2,\epsilon_3)$ is currently a sufficient space (and not including higher-order slow-roll parameters e.g. $\epsilon_4$) to characterise the single-field model selection capabilities of future CMB missions.} $\alphaS$ of the scalar spectral index in single-field inflation, defined as
\begin{equation}
\alphaS \equiv \left. \frac{\dd^2 \ln {\cal P}_{\zeta}}{\dd (\ln k)^2} \right\vert_{k_*}\,.
\end{equation}
In appendix~\ref{sec:alphaS-calculation} we derive a relation connecting the observed fiducial point and measurement width ($\muf^{\alphaS}$ and $\sigma^{\alphaS}$, respectively) over $\alphaS$ to the future widths over the slow-roll parameters $\boldsigma$, which we compute for each given realisation over the measured $\bmuf$ points. We shall not quote the relation here, but by referring to the functional dependencies $\muf^{\alphaS} = \muf^{\alphaS} (\bmuf , \boldsigma )$ and $\sigma^{\alphaS} = \sigma^{\alphaS} (\bmuf , \boldsigma )$ we can show that the probability which we seek is implicitly
\begin{align}\label{eq:alphaS-probability}
\Pr{}_{\alphaS > 2\sigma} (\boldsigma ) &\equiv \int_{\bmuf \in \mathbb{R}^n} p \left( \vert \muf^{\alphaS} \vert - 2 \sigma^{\alphaS} > 0 \, \vert \, \bmuf , \boldsigma \right) \dd \bmuf \\
&= \int_{\bmuf \in \mathbb{R}^n} \Theta \left[ \vert \muf^{\alphaS} (\bmuf , \boldsigma ) \vert - 2  \sigma^{\alphaS} (\bmuf , \boldsigma ) \right] \, p \left( \bmuf \vert {\cal D}_{\rm cur}\right) \dd \bmuf \label{eq:alphaS-probability-2}\,,
\end{align}
where we have specified a $2\sigma$-measurement over $\alphaS$ to be identified as having `measured $\alphaS$'.

In Table \ref{tab:alphaS-probs} we quote the probabilities of measurement over $\alphaS$ for each of the survey configurations studied in this paper. We find that for the survey P2 one obtains a substantial improvement over P1 in the probability of measuring $\alphaS$ --- moving from $\simeq 0.0$ to a probability of 0.93. When one reconsiders the posterior prediction, made this time when assuming that the Higgs Inflation model is `correct', we replace $p\left( \bmuf \vert {\cal D}_{\rm cur}\right)$ in \Eq{eq:alphaS-probability-2} with the posterior distribution $\bar{p}\left( \boldx \vert {\cal D}_{\rm cur},\calM_{{\rm HI}}\right)\propto \bar{\pi} \left( \boldx \vert \calM_{{\rm HI}} \right) \bar{\like}\left( {\cal D}_{\rm cur} \vert \boldx \right)$. From this change we see that there are significant probabilities for a detection of $\alphaS$ to be made by F2, F3 (and F4) surveys, hence improving the measurement over either $\epsilon_2$ or $\epsilon_3$ by an order of magnitude from the P2 survey. This can be seen explicitly through the relation in \Eq{eq:alphaS-width-formula}, where the otherwise relatively large term in the expression for $(\sigma^{\alphaS} )^2 \supset (\sigma^2)^2(\sigma^3)^2$ can only be reduced in size by decreasing either the measurement width over $\epsilon_2$ or $\epsilon_3$.

\begin{table}
\centering
\begin{tabular}{|c|c|c|c|}
\cline{1-4}
\multicolumn{2}{|c|}{Survey ($\boldsigma$)}  & $\Pr{}_{\alphaS > 2\sigma} (\boldsigma )$ & $\Pr{}_{\alphaS > 2\sigma} (\boldsigma )$ (HI posterior prediction) \\     \hline\hline
\cellcolor{yellow!55} & P1 & $0.0 + \varepsilon$ & $0.0 + \varepsilon$ \\      \hline
\cellcolor{red!55} & P2 & 0.93 & 0.02 \\      \hline
\cellcolor{blue!35} & F1 & 0.93 & 0.02  \\      \hline
\cellcolor{green!55} & F2 & 0.96 & 0.85 \\     \hline
\cellcolor{brown!90} & F3 & 0.96 & $1.0 - \varepsilon$ \\
\hline
\cellcolor{black!65} & F4 & 0.99 & $1.0 - \varepsilon$ \\   \hline
\end{tabular}
\caption{\label{tab:alphaS-probs} The probabilities of measurement over $\alphaS$ for each of the survey configurations studied in this paper, where measurement is defined as the fiducial point $\muf^{\alphaS}$ exceeding the $2\sigma$-error bound for a given future realisation. Note that $\varepsilon$ reminds the reader that the value is subject to rounding errors of up to $0.005$. In the final column we assume that HI is the `correct' model (replacing $p \, (\muf \vert {\cal D}_{\rm cur})$ with $\barp \, (\muf \vert {\cal D}_{\rm cur},\calM_{\rm HI} )$ in \Eq{eq:alphaS-probability-2}) and forecast the probability of detection of $\alphaS$ for each survey. }
\end{table}

\section{Concluding remarks}
\label{sec:concl}
In this paper we have outlined a simple method to compute any expected utility for a future survey given a previous set of measurements on the same variables from an independent survey. The tools that we have developed have all been included in \href{https://sites.google.com/view/foxicode}{\texttt{foxi}}, a publicly available python package that can be readily used in any survey forecasting problem. Crucially, our calculation relies on the assumption that the future likelihood can be modeled by an uncorrelated Gaussian distribution over the space of slow-roll parameters, hence, incorporating the level of detail required to tackle forecasting for proposed surveys like COrE/LiteBIRD must be an inevitable next step.

We have also modified the form of the expected utility in order to partition each possible future into either the rejection of models at the level of the maximum-likelihood or the decision between models using Bayesian model comparison. With the new expected utilities generated by this procedure, we have forecast the future of single-field inflationary model selection using 5 plateau potentials that are both indicative of the class and span the range of observables $(\epsilon_1,\epsilon_2,\epsilon_3)$ --- the slow-roll parameters --- that is typical for models of this type (see appendix~\ref{sec:models} for their definitions). Our analysis finds two important thresholds in the space of missions:
\begin{enumerate}
\item{Increasing precision from a P1-type survey capabilities (like CMB Stage-4) to P2 (like LiteBIRD/COrE), we cross the first threshold where most of the possible future measurements that could be made will rule out both single-field models of each pair at the level of their maximum likelihoods. }
\item{Increasing precision from F1-3 to F4-type toy survey capabilities, we cross a second threshold where our utility functions saturate to constant values that do not depend on the precision of the measurement. In this limit, the widths of the future likelihoods are much smaller than the prior volumes from the models that we consider. For both models of a given pair not to be rejected at the level of the maximum likelihood, the value of $\bmuf$ must fall within at least one of their prior volumes. If this is so then the Bayes factor becomes the ratio between their prior densities at that point (see \Eq{eq:trivial-bayes-factor}) which does not depend on the future measurement widths.}
\end{enumerate}

The prior volume-dominated limit, arising from threshold 2 above, is analogous to the threshold reached within our computational procedure (outlined in appendix~\ref{sec:foxi-computation}), where in the latter case we devise a method to calculate the Bayesian evidence that relies upon \Eq{eq:trivial-bayes-factor}. Once the threshold of this regime has been crossed it is \emph{essential} for more theoretical progress in the understanding of the remaining models to occur, which would result in more narrow priors on their parameters, before one builds a new survey to choose between them

Though the space of surveys that we explore in this work may be simplistic, the broad conclusions we draw are unlikely to change. Our results using only information theory considerations (the expected Kullback-Leibler divergence $\langle \dkl \rangle$) indicat1e that the greatest information to be gained is on $\epsilon_3$, since it is currently the least constrained of the three slow-roll parameters (and may also be used to detect a scalar running). However, our analysis also suggests that the most-likely decisive gains in selecting between single-field inflationary models are made by improving the second slow-roll parameter $\epsilon_2$ constraint (which can also potentially be used to detect a scalar running) --- which can be measured through more precision on the scalar spectral index $\nS$. Finally, as is suggested by many theoretical studies into the fundamental physics of quantum gravity, the tensor-to-scalar ratio $r$ might be the most important CMB observable and hence $\epsilon_1$ may be considered the most fundamentally attractive to theorists. Therefore, to order this trichotomy, we have compiled the following list:
\begin{enumerate}
\item{Improve the measurement over $\nS$, hence $\epsilon_2$ will be constrained to a greater degree and therefore one optimises the single-field slow-roll decisivity. Also we may potentially observe $\alphaS$.}
\item{Improve the measurement over $r$, hence $\epsilon_1$ will be constrained to a greater degree and we may learn more about fundamental physics.}
\item{Improve the measurement over $\alphaS$, hence $\epsilon_3$ will be constrained to a greater degree which is optimal from an information-theoretic standpoint.}
\end{enumerate}

We also considered the applications of our framework to forecasting the potential of surveys to infer the temperature of reheating, given the Higgs inflationary potential. This is an avenue which we only very briefly have explored in this work but a clear extension would be to conduct a more thorough analysis on reheating temperatures taking into account different choices of inflationary potentials that still match observations. This also serves to illustrate the next step in the challenges set to model-builders in the future: one must be more specific in predicting reheating temperatures that arise from a given inflationary potential as one approaches the second threshold.

In \Sec{sec:measuring-alphaS} we have promoted an additional application of our framework to obtaining probabilities of measuring a given parameter in the future. In this case, we considered the probability of measuring the scalar running $\alphaS$, initially when assuming no preferred model, and then subsequently when assuming that a slow-roll single-field model (the HI model in this case) is preferred and hence the current data is the posterior prediction of the model from \emph{Planck}. Our results broadly indicate that though a P2-like survey is generally expected to measure $\alphaS$, if the \emph{Planck} posterior is consistent with a slow-roll single-field model then the probability of such a measurement drops dramatically and it is only with more advanced mock surveys like F2 or F3 that the chances of measuring $\alphaS$ become significant once again. This can be traced to the fact that $\alphaS$ is typically small to be consistent with slow-roll single-field models, and hence a more advanced survey is required to measure its potential deviation away from 0.

We have explored many ideas in this first concrete step into the new territory of Bayesian experimental design for model selection in the context of cosmological experiments. We will conclude with a list of some extensions that can be made to this work:
\begin{enumerate}
\item{The analysis here can be performed for a specific survey by specifying more detail in the functional form of ${\cal D}_{\rm fut}$ in \Eq{eq:exu} that includes detector behaviour. }
\item{Optimising $\boldsigma$ with respect to a given expected utility, e.g. the decisivity, in a series of sequential surveys can define a path in the future over the $\boldsigma$ space. A formalism such as that of Information Geometry~\cite{amari2009information, Brehmer:2016nyr} could prove very useful to this end.}
\item{One could use our method to optimise the connection between scientific gains and their required financial inputs. In other words, find the optimal point in the space of $\boldsigma$ under the constraint of a given budget and translate this into specifications of a survey (e.g. number of detectors, frequency channels, noise sensitivity, angular resolution, telescope size, etc...).}
\item{It is natural to also consider an identical analysis for beyond single-field models of inflation, e.g. curvaton models and other multi-field scenarios as there are already numerical implementations of these model priors~\cite{Price:2014xpa,Vennin:2015vfa} which represent minimal extensions to single-field inflation that can still be consistent with the current data.}
\end{enumerate}

\section*{Acknowledgments}
We would like to thank Paul Carter, Thomas Collett, Florent Leclercq and Jes\'{u}s Torrado for engaging discussions with enlightening comments. RJH is supported by UK Science and Technology Facilities
Council grant ST/N5044245. VV acknowledges funding from the European Union’s Horizon 2020 research and innovation programme under the Marie Sklodowska-Curie grant agreement ${\rm N}^0$ 750491. VV and DW acknowledge financial support from UK Science and Technology Facilities Council grant ST/N000668/1. Some numerical computations were done on the Sciama High Performance Compute (HPC) cluster which is supported by the ICG, SEPNet and the University of Portsmouth.
\begin{landscape}
\begin{table}
\centering
\resizebox{0.45\columnwidth}{!}{%
\begin{tabular}{|c|c|c|c|c|c|}
\cline{2-6}
\multicolumn{1}{c}{\cellcolor{blue!35}} & \multicolumn{5}{|c|}{F1 with $\pi (\bmuf \vert \log \epsilon_1)$}  \\      \hline
\backslashbox{$\calM_\beta$ - $\calM_\gamma$}{$\langle  U\rangle$}  & $\langle \vert \ln {\rm B}_{\beta \gamma}\vert \rangle$ & $\langle \vert \ln {\rm B}_{\beta \gamma} \vert \rangle_{{}_{\rm ML}}$ & $\deci_{\beta \gamma}$ & $\deci_{\beta \gamma}\vert_{{}_{\rm ML}}$ & $r_{{}_\mathrm{ML}}$ \\     \hline\hline
${\rm KMIII}$ - ${\rm HI}$ & $261.90 \,(\pm 110.24)$ & $295.11 \,(\pm 155.11)$ & 0.96 & $1.0-\varepsilon$ & 0.95  \\      \hline
${\rm KKLTI}_{\rm stg}$ - ${\rm HI}$ & $262.36 \,(\pm 104.38)$ & $262.26 \,(\pm 37.01)$ & 0.96 & $1.0-\varepsilon$ & 0.95  \\      \hline
${\rm LI}_{\alpha >0}$ - ${\rm HI}$ & $249.77 \,(\pm 99.15)$ & $242.08 \,(\pm 133.01)$ & 0.96 & 0.96 & 0.97  \\      \hline
${\rm RGI}$ - ${\rm HI}$ & $270.64 \,(\pm 133.67)$ & $277.18 \,(\pm 184.55)$ & 0.98 & 0.97 & 0.95  \\      \hline
${\rm KKLTI}_{\rm stg}$ - ${\rm KMIII}$ & $5.01 \,(< 97.20)$ & $41.17 \,(< 389.58)$ & 0.08 & 0.08 & 0.95  \\      \hline
${\rm LI}_{\alpha >0}$ - ${\rm KMIII}$ & $38.09 \,(< 323.48)$ & $89.15 \,(< 521.72)$ & 0.85 & 0.70 & 0.95  \\      \hline
${\rm RGI}$ - ${\rm KMIII}$ & $50.84 \,(< 408.62)$ & $123.43 \,(< 624.76)$ & 0.11 & 0.23 & 0.94  \\      \hline
${\rm LI}_{\alpha >0}$ - ${\rm KKLTI}_{\rm stg}$ & $37.81 \,(< 309.16)$ & $47.81 \,(< 364.24)$ & 0.91 & 0.60 & 0.95  \\      \hline
${\rm RGI}$ - ${\rm KKLTI}_{\rm stg}$ & $50.54 \,(< 397.36)$ & $84.83 \,(< 519.96)$ & 0.16 & 0.18 & 0.94  \\      \hline
${\rm RGI}$ - ${\rm LI}_{\alpha >0}$ & $21.77 \,(< 257.82)$ & $36.40 \,(< 353.30)$ & 0.40 & 0.12 & 0.95  \\      \hline
\end{tabular}}
\resizebox{0.45\columnwidth}{!}{%
\begin{tabular}{|c|c|c|c|c|c|}
\cline{2-6}
\multicolumn{1}{c}{\cellcolor{green!55}} & \multicolumn{5}{|c|}{F2 with $\pi (\bmuf \vert \log \epsilon_1)$}   \\      \hline
\backslashbox{$\calM_\beta$ - $\calM_\gamma$}{$\langle  U\rangle$}  & $\langle \vert \ln {\rm B}_{\beta \gamma}\vert \rangle$ & $\langle \vert \ln {\rm B}_{\beta \gamma} \vert \rangle_{{}_{\rm ML}}$ & $\deci_{\beta \gamma}$ & $\deci_{\beta \gamma}\vert_{{}_{\rm ML}}$ & $r_{{}_\mathrm{ML}}$ \\  \hline\hline
${\rm KMIII}$ - ${\rm HI}$ & $28.11 \,(< 172.58)$ & $71.59 \,(< 490.44)$ & 0.86 & 0.47 & 0.97  \\      \hline
${\rm KKLTI}_{\rm stg}$ - ${\rm HI}$ & $18.18 \,(< 149.76)$ & $51.05 \,(< 204.78)$ & 0.34 & 0.50 & 0.96  \\      \hline
${\rm LI}_{\alpha >0}$ - ${\rm HI}$ & $275.01 \,(\pm 172.90)$ & $96.64 \,(\pm 40.41)$ & 0.97 & $1.0-\varepsilon$ & 0.99  \\      \hline
${\rm RGI}$ - ${\rm HI}$ & $77.96 \,(< 254.88)$ & $26.79 \,(< 219.88)$ & 0.93 & 0.50 & 0.98  \\      \hline
${\rm KKLTI}_{\rm stg}$ - ${\rm KMIII}$ & $40.43 \,(< 222.38)$ & $86.58 \,(< 425.10)$ & 0.80 & 0.58 & 0.95  \\      \hline
${\rm LI}_{\alpha >0}$ - ${\rm KMIII}$ & $298.15 \,(\pm 185.73)$ & $210.27 \,(< 425.30)$ & 0.98 & $1.0-\varepsilon$ & 0.97  \\      \hline
${\rm RGI}$ - ${\rm KMIII}$ & $104.06 \,(< 303.76)$ & $86.86 \,(< 457.74)$ & 0.95 & 0.79 & 0.96  \\      \hline
${\rm LI}_{\alpha >0}$ - ${\rm KKLTI}_{\rm stg}$ & $266.92 \,(\pm 174.48)$ & $160.60 \,(< 385.16)$ & 0.96 & 0.90 & 0.96  \\      \hline
${\rm RGI}$ - ${\rm KKLTI}_{\rm stg}$ & $75.78 \,(< 246.76)$ & $63.35 \,(< 253.52)$ & 0.90 & 0.63 & 0.95  \\      \hline
${\rm RGI}$ - ${\rm LI}_{\alpha >0}$ & $217.27 \,(\pm 158.11)$ & $65.05 \,(< 189.44)$ & 0.97 & 0.94 & 0.98  \\      \hline
\end{tabular}}
\resizebox{0.45\columnwidth}{!}{%
\begin{tabular}{|c|c|c|c|c|c|}
\cline{2-6}
\multicolumn{1}{c}{\cellcolor{brown!90}} & \multicolumn{5}{|c|}{F3 with $\pi (\bmuf \vert \log \epsilon_1)$}   \\      \hline
\backslashbox{$\calM_\beta$ - $\calM_\gamma$}{$\langle  U\rangle$}  & $\langle \vert \ln {\rm B}_{\beta \gamma}\vert \rangle$ & $\langle \vert \ln {\rm B}_{\beta \gamma} \vert \rangle_{{}_{\rm ML}}$ & $\deci_{\beta \gamma}$ & $\deci_{\beta \gamma}\vert_{{}_{\rm ML}}$ & $r_{{}_\mathrm{ML}}$ \\     \hline\hline
${\rm KMIII}$ - ${\rm HI}$ & $5.64 \,(< 127.44)$ & $232.18 \,(< 834.90)$ & - & 0.58 & 0.99  \\      \hline
${\rm KKLTI}_{\rm stg}$ - ${\rm HI}$ & $25.68 \,(< 268.98)$ & $92.59 \,(< 216.98)$ & - & 0.83 & 0.97  \\      \hline
${\rm LI}_{\alpha >0}$ - ${\rm HI}$ & $4.41 \,(< 100.92)$ & $4.06 \,(\pm 3.81)$ & - & 0.31 & 0.99  \\      \hline
${\rm RGI}$ - ${\rm HI}$ & $6.28 \,(< 130.06)$ & $22.58 \,(< 226.72)$ & - & 0.37 & 0.99  \\      \hline
${\rm KKLTI}_{\rm stg}$ - ${\rm KMIII}$ & $24.75 \,(< 270.18)$ & $141.92 \,(< 474.28)$ & - & 0.80 & 0.97  \\      \hline
${\rm LI}_{\alpha >0}$ - ${\rm KMIII}$ & $3.81 \,(< 109.54)$ & $218.19 \,(< 817.54)$ & - & 0.45 & 0.99  \\      \hline
${\rm RGI}$ - ${\rm KMIII}$ & $5.12 \,(< 132.58)$ & $222.51 \,(< 812.02)$ & - & 0.30 & 0.99  \\      \hline
${\rm LI}_{\alpha >0}$ - ${\rm KKLTI}_{\rm stg}$ & $23.74 \,(< 261.16)$ & $88.34 \,(< 219.20)$ & - & 0.79 & 0.96  \\      \hline
${\rm RGI}$ - ${\rm KKLTI}_{\rm stg}$ & $24.88 \,(< 270.06)$ & $91.21 \,(< 243.78)$ & - & 0.76 & 0.96  \\      \hline
${\rm RGI}$ - ${\rm LI}_{\alpha >0}$ & $2.14 \,(< 83.62)$ & $15.23 \,(< 196.92)$ & - & 0.07 & 0.99  \\      \hline
\end{tabular}}
\begin{adjustbox}{totalheight=0.3\textheight,scale={1.0}{1.035}}
\begin{tabular}{|c|c|c|c|c|c|}
\cline{2-6}
\multicolumn{1}{c}{\cellcolor{black!65}} & \multicolumn{5}{|c|}{F4 with $\pi (\bmuf \vert \log \epsilon_1)$}  \\      \hline
\backslashbox{$\calM_\beta$ - $\calM_\gamma$}{$\langle  U\rangle$}  & $\langle \vert \ln {\rm B}_{\beta \gamma}\vert \rangle$ & $\langle \vert \ln {\rm B}_{\beta \gamma} \vert \rangle_{{}_{\rm ML}}$ & $\deci_{\beta \gamma}$ & $\deci_{\beta \gamma}\vert_{{}_{\rm ML}}$ & $r_{{}_\mathrm{ML}}$ \\     \hline\hline
${\rm KMIII}$ - ${\rm HI}$ & $36.00 \,(< 324.80)$ & $541.00 \,(\pm 334.40)$ & - & $1.0-\varepsilon$ & $1.0-\varepsilon$  \\      \hline
${\rm KKLTI}_{\rm stg}$ - ${\rm HI}$ & $43.24 \,(< 357.98)$ & $481.08 \,(\pm 184.16)$ & - & $1.0-\varepsilon$ & 0.97  \\      \hline
${\rm LI}_{\alpha >0}$ - ${\rm HI}$ & $22.82 \,(< 257.78)$ & $354.80 \,(\pm 140.62)$ & - & $1.0-\varepsilon$ & $1.0-\varepsilon$  \\      \hline
${\rm RGI}$ - ${\rm HI}$ & $34.43 \,(< 315.94)$ & $331.07 \,(\pm 224.09)$ & - & $1.0-\varepsilon$ & $1.0-\varepsilon$  \\      \hline
${\rm KKLTI}_{\rm stg}$ - ${\rm KMIII}$ & $15.58 \,(< 198.24)$ & $195.91 \,(< 449.30)$ & - & 0.98 & 0.97  \\      \hline
${\rm LI}_{\alpha >0}$ - ${\rm KMIII}$ & $29.18 \,(< 294.32)$ & $437.89 \,(\pm 386.77)$ & - & $1.0-\varepsilon$ & $1.0-\varepsilon$  \\      \hline
${\rm RGI}$ - ${\rm KMIII}$ & $18.94 \,(< 238.12)$ & $299.98 \,(< 847.90)$ & - & 0.91 & 0.99  \\      \hline
${\rm LI}_{\alpha >0}$ - ${\rm KKLTI}_{\rm stg}$ & $31.98 \,(< 300.82)$ & $307.15 \,(\pm 220.99)$ & - & $1.0-\varepsilon$ & 0.97  \\      \hline
${\rm RGI}$ - ${\rm KKLTI}_{\rm stg}$ & $20.93 \,(< 240.40)$ & $191.94 \,(\pm 183.46)$ & - & 0.97 & 0.97  \\      \hline
${\rm RGI}$ - ${\rm LI}_{\alpha >0}$ & $19.99 \,(< 231.98)$ & $131.58 \,(< 420.74)$ & - & 0.98 & $1.0-\varepsilon$  \\      \hline
\end{tabular}
\end{adjustbox}
\caption{~\label{tab:f14-utilities} Computed expected utilities for the F1-4 experiments in the case where the $\pi (\bmuf \vert \log \epsilon_1)$ prior is used (see \Eq{eq:predictive-priors-logeps1}). Note that $\varepsilon$ reminds the reader that the value is subject to rounding errors of up to $0.005$. Values in brackets $\pm$ around each computed expected utility correspond to the $1$-$\sigma$ errors, which are evaluated using \Eq{eq:secondmomentu}. This symmetric error about our different expected values for $\vert \ln {\rm B}_{\beta \gamma} \vert$ is replaced with a $2$-$\sigma$ upper bound (because it is positive by definition) if the lower error is greater than the expected value itself.}
\end{table}
\end{landscape}
\newpage
\appendix
\section{The single-field models} \label{sec:models}

The observational predictions from each of the models defined below have all been calculated using the publicly available \texttt{ASPIC} library: \href{http://cp3.irmp.ucl.ac.be/~ringeval/aspic.html}{http://cp3.irmp.ucl.ac.be/~ringeval/aspic.html}. The model priors were obtained from \Ref{Martin:2013nzq} and we have also provided arguments for the choice of each model as representatives of the full sample.

\textbf{Higgs Inflation (HI)} has the following potential
\begin{equation} \label{eq:higgs-pot}
\ V = M^4 \left[ 1-\exp \left( -\sqrt{\frac{2}{3}}\frac{\phi}{\Mp}\right) \right] \,,
\end{equation}
and was chosen in our analysis of plateaus to represent models with a relatively large tensor-to-scalar ratio. In addition, the fact that it is effectively a 0-free-parameter model is attractive with respect to Bayesian inference.

\textbf{Loop Inflation (${\rm LI}_{\alpha >0}$)} with a particular prior choice for the $\alpha$ parameter
\begin{equation}
\ V = M^4 \left[ 1 + \alpha \ln \left( \frac{\phi}{\Mp} \right) \right] \,, \qquad \log (\alpha ) \in [\log (0.003), \log (0.3) ]\,,
\end{equation}
was considered here for its relatively large spectral index, thus ideally providing a decisive tension with the HI and KMIII models in particular.

\textbf{Radion Gauge Inflation (RGI)} was chosen with the following potential and prior
\begin{equation}
\ V = M^4 \frac{(\phi /\Mp )^2}{\alpha + (\phi /\Mp )^2}\,, \qquad \log (\alpha ) \in [-4,4] \,,
\end{equation}
and is a good all-round representative of a standard plateau model that is favoured by observations with a reasonably large tensor-to-scalar ratio. The model is also in a good position between HI and ${\rm LI}_{\alpha >0}$ in values of the spectral index.

\textbf{K\"{a}hler Moduli Inflation II (KMIII)} is a good example of a two-parameter plateau model with the following potential and choices of parameters
\begin{equation}
\ V = M^4 \left[ 1 - \alpha \frac{\phi}{\Mp}\exp \left( -\beta \frac{\phi}{\Mp}\right) \right] \,, \qquad \log ({\cal V}) \in [5,7]\,, \qquad \frac{\alpha}{\beta {\cal V}} \in [0.2,5]\,,
\end{equation}
where one calculates $\beta = {\cal V}^{2/3}$ and sets $\alpha$ through the ratio $\alpha /(\beta {\cal V})$. This model also has a much lower order of magnitude for the tensor-to-scalar ratio in comparison with the three above, mapping out a more complete region of the $(\nS ,r)$-diagram.

\textbf{Kachru-Kallosh-Linde-Trivedi Inflation (${\rm KKLTI}_{\rm stg}$)} phenomenologically interpolates between much of the currently available parameter space with the potential and the following potential and priors
\begin{equation}
\ V = \frac{M^4}{1+\left( \frac{\mu}{\phi} \right)^{4}} \,, \qquad \log \left( \frac{\mu}{\Mp}\right) \in [-6,\log (2)]\,,
\end{equation}
thus it is a good final addition to our small sample of models.

A summary plot of the available parameter space on the $(\nS , r)$-diagram for each of the models is shown in \Fig{fig:nsr-plot}, where it is immediately clear that we have selected a reasonable sample of single-field models to span the available parameter space.

\section{Computational methods in \texttt{foxi}}
\label{sec:foxi-computation}

\begin{figure}[t]
\begin{center}
\includegraphics[width=0.85\textwidth]{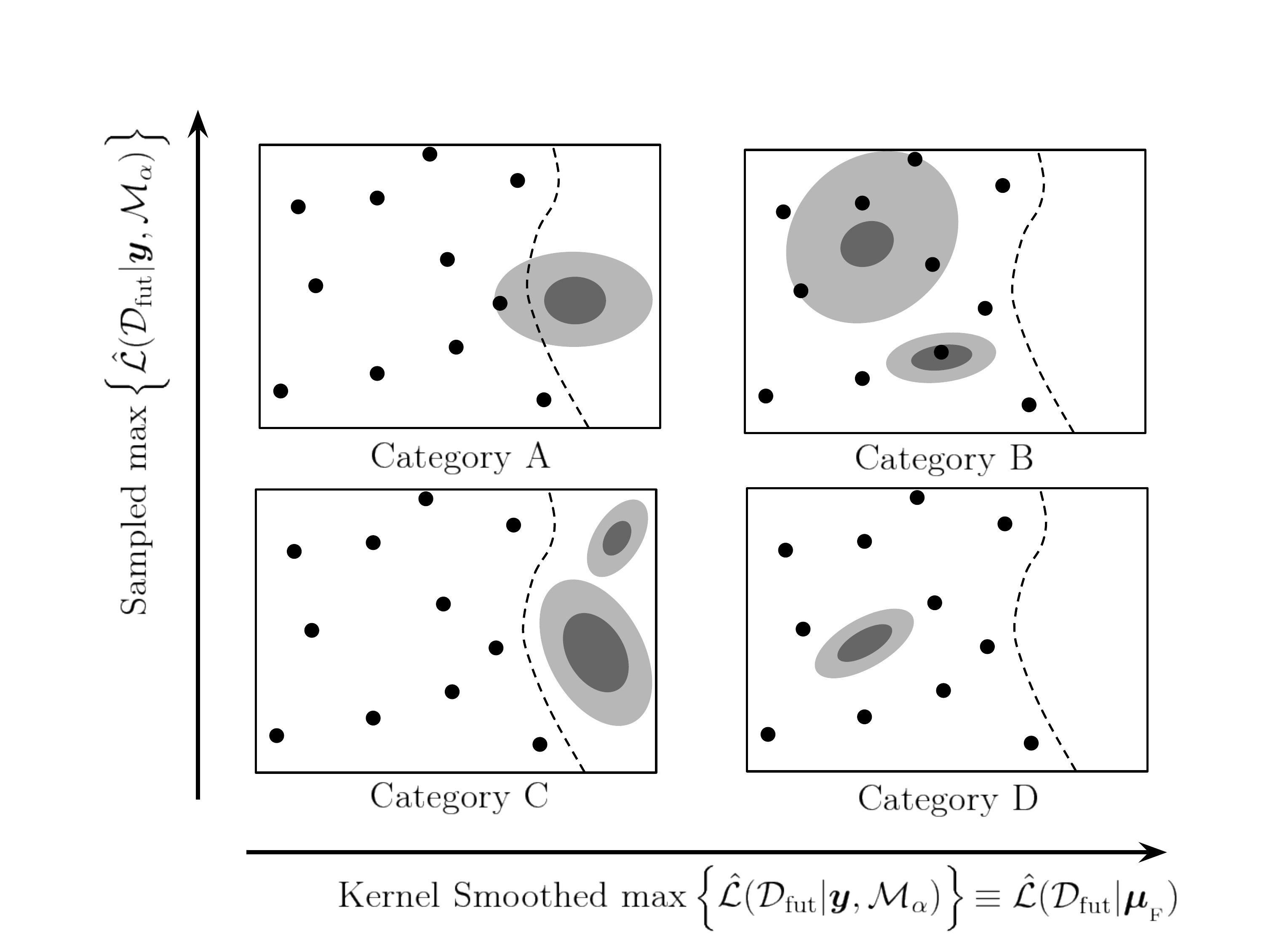}
\caption{~\label{fig:foxi-diagram-categories} A diagram depicting 4 unique categories of scenario practically encountered in the computation of the Bayesian evidence using the approximation \Eq{eq:lnB-approx-discrete}. The black dots signify the prior chain samples, the shaded contours are the $1$-$\sigma$ and $2$-$\sigma$ limits of the future likelihood modeled with a Gaussian and the region to the left of the dotted curved line in all 4 boxes indicates the outer contour of the kernel-smoothed prior density using the samples and \Eq{eq:kde}. Boxes further to the right have larger maximum likelihood values contained within the prior obtained from kernel smoothing and boxes further upward have larger maximum likelihood values using the prior samples directly. Category A arises from only a mild overlap between the kernel-smoothed density and the future likelihood contour. Category B denotes either the future likelihood contour is quite large or is small but serendipitously centred directly over a $\bmuf$ point. Category C situations produce Bayesian evidences that are rightfully considered to be always ruled out beyond the Jeffrey's threshold. Category D situations have a very small future likelihood contour --- below the typical inter-point distance of the prior samples.  }
\end{center}
\end{figure}

In \Fig{fig:foxi-diagram-categories} we provide a reference diagram illustrating the various situations which arise during computation of the utility functions in the main body of work. In particular, the Bayesian evidence approximation of \Eq{eq:lnB-approx-discrete}  practically requires the integration over the probability densities described by both a Gaussian function and prior samples. These distributions can be easily combined when the future likelihood described by the Gaussian function has relatively wide $1$-$\sigma$ contour limits compared to the typical inter-point distance of the prior chains --- such as is true for the category A situations depicted in \Fig{fig:foxi-diagram-categories} and some situations within category B.

Category D (and category B points with a relatively small error contour) represent situations where we must adopt a different computational approach. A convenient non-parametric method is to approximate the model prior probability density $\barpi (\boldx \vert \calM_\alpha )$ using Kernel Density Estimation
\begin{equation} \label{eq:kde}
\barpi (\boldx \vert \calM_\alpha ) \simeq \frac{1}{Z_\alpha} \sum_{\boldx_i \in \{ \calM_\alpha \, {\rm chains} \}} {\cal K}_{\boldsymbol{w}} (\boldx ,\boldx_i)\,,
\end{equation}
or `kernel smoothing', as illustrated in the right-hand column of boxes in \Fig{fig:foxi-diagram-categories}. $Z_\alpha$ in \Eq{eq:kde} is simply the number of samples within the Markov chains representing the prior of $\calM_\alpha$. In this work, the Kernel ${\cal K}_{\boldsymbol{w}}$ we select is simply a Gaussian function
\begin{equation}
\ {\cal K}_{\boldsymbol{w}}(\boldsymbol{a},\boldsymbol{b}) = (2\pi )^{-\frac{n}{2}} \left( \prod^n_{i=1} w^i \right)^{-1} \exp \left[ - \sum^n_{i=1}\frac{(a^i-b^i)^2}{2(w^i)^2} \right] \,,
\end{equation}
with bandwidth vector $\boldsymbol{w}$. Though Category D situations are easily identifiable because the maximum likelihood obtained from direct samples is much lower than the kernel-smoothed equivalent, in general, we have to use an optimal estimate~\footnote{In our case we use the in-built Least-Squares Cross-Validation (LSCV) method implemented in the \href{http://www.statsmodels.org/dev/install.html}{\texttt{statsmodels}} package in python. LSCV is based on minimising the integrated square error between the estimated distribution $f_{\rm est}\propto \sum {\cal K}_{\boldsymbol{w}}$ and the underlying true distribution $f_{\rm true}$ i.e. minimising
\begin{equation}
\int_{\boldz \in \mathtt{R}^n} \left[ \frac{1}{M}\sum_{\boldz_i \in \left\{ {\rm Samples} \right\} } {\cal K}_{\boldsymbol{w}}(\boldz ,\boldz_i) - f_{\rm true}(\boldz )\right]^2 \dd \boldz \,,
\end{equation}
with $M$ samples, by minimising Silverman's~\cite{1986desd.book.....S} estimator
\begin{equation}
\ S = \int_{\boldz \in \mathtt{R}^n} \frac{1}{M^2} \left[ \sum_{\boldz_i \in \left\{ {\rm Samples} \right\} } {\cal K}_{\boldsymbol{w}}(\boldz ,\boldz_i) \right]^2 \dd \boldz - \frac{2}{M} \sum_{\boldz_j \in \{ {\rm Samples}\}} \sum_{ \forall \boldz_i\neq \boldz_j} {\cal K}_{\boldsymbol{w}}(\boldz_j ,\boldz_i) \,.
\end{equation}} of $\boldsymbol{w}$ to identify whether kernel smoothing is necessary in Category B i.e. if we are in regions where the local density of points is too sparse, we will find that one or more of the dimensions within $\boldsymbol{w}$ will fall outside the corresponding dimension of the $1$-$\sigma$ futuristic likelihood contour.

In the limit where the futuristic likelihood contour is very small compared with the typical $\boldsymbol{w}$ one finds for the smoothed prior chains, to good approximation we find that the local value $\barpi (\boldx \vert \calM_\alpha ) \propto {\rm const.}$ and therefore we need only compute the evidence (and the maximum likelihood point) using a single prior value centred at the $\bmuf$ point
\begin{equation} \label{eq:evidence-kde-category-b-d}
\left. {\cal E}_\alpha \, \right\vert_{\boldsigma \ll  \boldsymbol{w}} \simeq \barpi (\bmuf \vert \calM_\alpha ) \simeq \frac{1}{Z_\alpha} \sum_{\boldx_i \in \{ \calM_\alpha \, {\rm chains} \}} {\cal K}_{\boldsymbol{w}} (\bmuf ,\boldx_i) \,.
\end{equation}
Though this estimate can be shown to be very accurate, the \href{https://sites.google.com/view/foxicode}{\texttt{foxi}} algorithm itself actually computes the Bayesian evidence in the regime of some category B and all category D situations by implementing the combined approach of both \Eq{eq:evidence-kde-category-b-d} and drawing typically 1000 samples from the future likelihood (\Eq{eq:future-gaussian-assumption}) to sum over for the integral. This method is more computationally robust than \Eq{eq:evidence-kde-category-b-d} alone since it can accommodate for scenarios where the magnitudes of error in each dimension in $\boldsigma$ are very different, offering greater flexibility to the algorithm, at a cost of some additional computation time and efficiency.

\begin{table}
\centering
\begin{tabular}{|c|c|c|c|c|c|}
\cline{1-6}
\multicolumn{2}{|c|}{Survey}  & {\rm Ave. Category A} & {\rm Ave. Category B} & {\rm Ave. Category C} & {\rm Ave. Category D} \\     \hline\hline
\cellcolor{yellow!55} & P1 & 84.1 \% & $0.0+\varepsilon$ \% & 21.9 \% & $0.0 +\varepsilon$ \% \\      \hline
\cellcolor{red!55} & P2 & 3.7 \% & 0.8 \% & 95.5 \% & $0.0 +\varepsilon$ \%  \\      \hline
\cellcolor{blue!35} & F1 & 2.8 \% & 0.8 \% & 96.4 \% & $0.0 +\varepsilon$ \%  \\      \hline
\cellcolor{green!55} & F2 & 1.3 \% & 0.6 \% & 97.9 \% & 0.2 \%  \\     \hline
\cellcolor{brown!90} & F3 & 0.5 \% & 0.4 \% & 98.8 \% & 0.3 \%  \\
\hline
\cellcolor{black!65} & F4 & $0.0 +\varepsilon$ \% & 0.1 \% & 99.3 \% & 0.6 \%  \\   \hline
\end{tabular}
\caption{\label{tab:percentage-points-categories} The percentage number of $\bmuf$ points in the Markov chains representing the \emph{Planck} data that correspond to the computational situations defined in \Fig{fig:foxi-diagram-categories}. Note that $\varepsilon$ reminds the reader that the value is subject to rounding errors of up to $0.005$. }
\end{table}

\section{Checking for numerical robustness} \label{sec:gaussian-assumption-limitations}

This section aims to quantify empirically the accuracy of the Gaussian assumption used throughout this work with respect to the direct applicability of our mock forecasts to `real-world' surveys. Note that we are not suggesting that the assumption is `incorrect' in any sense, but that by definition, forecasting using the Gaussian assumption does not necessarily coincide with a true likelihood that would be obtained from a specific survey forecast.

We compared our results for each model pair using \Eq{eq:lnB-approx-discrete} with those obtained from the \href{http://www.mrao.cam.ac.uk/software/multinest/}{\texttt{MultiNest}}~\cite{Feroz:2008xx, Martin:2013nzq} algorithm in each case, where we obtained both $\bmuf$ and $\boldsigma$ for \Eq{eq:lnB-approx-discrete} through the prior samples and a Gaussian likelihood with mean and marginalised variances computed from the chains\footnote{The specifications used to forecast the likelihood for LiteCOrE are given in \Ref{Finelli:2016cyd} and correspond to what is referred to as `LiteCORE-120'.} used by \texttt{MultiNest}, respectively. A comparison is in Table \ref{tab:planck-foxi-comparison} for the \emph{Planck} 2015 data~\cite{Ade:2015oja}, where there is good general agreement up to the $\ln {\rm B}_{\beta \gamma} \pm 0.6$ level, and the forecast data for the LiteCOrE forecast dataset~\cite{DiValentino:2016foa,Finelli:2016cyd} using HI fixed with $T_{\rm reh}=10^6{\rm GeV}$ as the fiducial model, where there is less consistent agreement up to the $\ln {\rm B}_{\beta \gamma} \pm 5.0$ level.

\begin{table}
\centering
\begin{tabular}{|c|c|c|c|c|}
\cline{1-5}
 & \multicolumn{2}{c|}{\emph{Planck} 2015} & \multicolumn{2}{c|}{LiteCOrE (HI fiducial)}  \\      \hline
\backslashbox{$\calM_\beta$ - $\calM_\gamma$}{$\ln {\rm B}_{\beta \gamma}$}  & Gaussian & \href{ http://www.mrao.cam.ac.uk/software/multinest/}{\texttt{MultiNest}}~\cite{Feroz:2008xx, Martin:2013nzq} & Gaussian & \href{ http://www.mrao.cam.ac.uk/software/multinest/}{\texttt{MultiNest}} \\ \hline\hline
${\rm KMIII}$ - ${\rm HI}$ & 0.11 & 0.04 & -3.52 & -7.63   \\      \hline
${\rm KKLTI}_{\rm stg}$ - ${\rm HI}$ & -0.57 & -0.44 & -3.66 & -8.02    \\      \hline
${\rm LI}_{\alpha > 0}$ - ${\rm HI}$ & -2.33 & -2.48 & -18.11 & -17.89  \\      \hline
${\rm RGI}$ - ${\rm HI}$ & -0.92 & -0.68 & -4.32 & -4.63   \\      \hline
${\rm KKLTI}_{{\rm stg}}$ - ${\rm KMIII}$ & -0.68 & -0.48 & -0.14 & -0.39   \\      \hline
${\rm LI}_{\alpha > 0}$ - ${\rm KMIII}$ & -2.44 & -2.51 & -14.60 & -10.25  \\      \hline
${\rm RGI}$ - ${\rm KMIII}$ & -1.03 & -0.71 & -0.80 & 3.00  \\ \hline
${\rm LI}_{\alpha > 0}$ - ${\rm KKLTI}_{{\rm stg}}$ & -1.76 & -2.04 &  -14.45 & -9.86 \\      \hline
${\rm RGI}$ - ${\rm KKLTI}_{{\rm stg}}$ & -0.35 & -0.23 & -0.67 & 3.39  \\      \hline
${\rm RGI}$ - ${\rm L7I}_{\alpha > 0}$ & 1.41 & 1.81 & 13.79 & 13.25  \\      \hline
\end{tabular}
\caption{~\label{tab:planck-foxi-comparison} A comparison table showing the differences between Bayes factors approximated with a Gaussian assumption (denoted `Gaussian') to those obtained from the \href{ http://www.mrao.cam.ac.uk/software/multinest/}{\texttt{MultiNest}}~\cite{Feroz:2008xx, Martin:2013nzq, Ringeval:2013lea} algorithm in each case of model pair for the \emph{Planck} 2015 and forecast LiteCOrE~\cite{DiValentino:2016foa,Finelli:2016cyd} (with Higgs Inflation as a fiducial model) datasets. }
\end{table}

\begin{figure}[t]
\begin{center}
\includegraphics[width=0.6\textwidth]{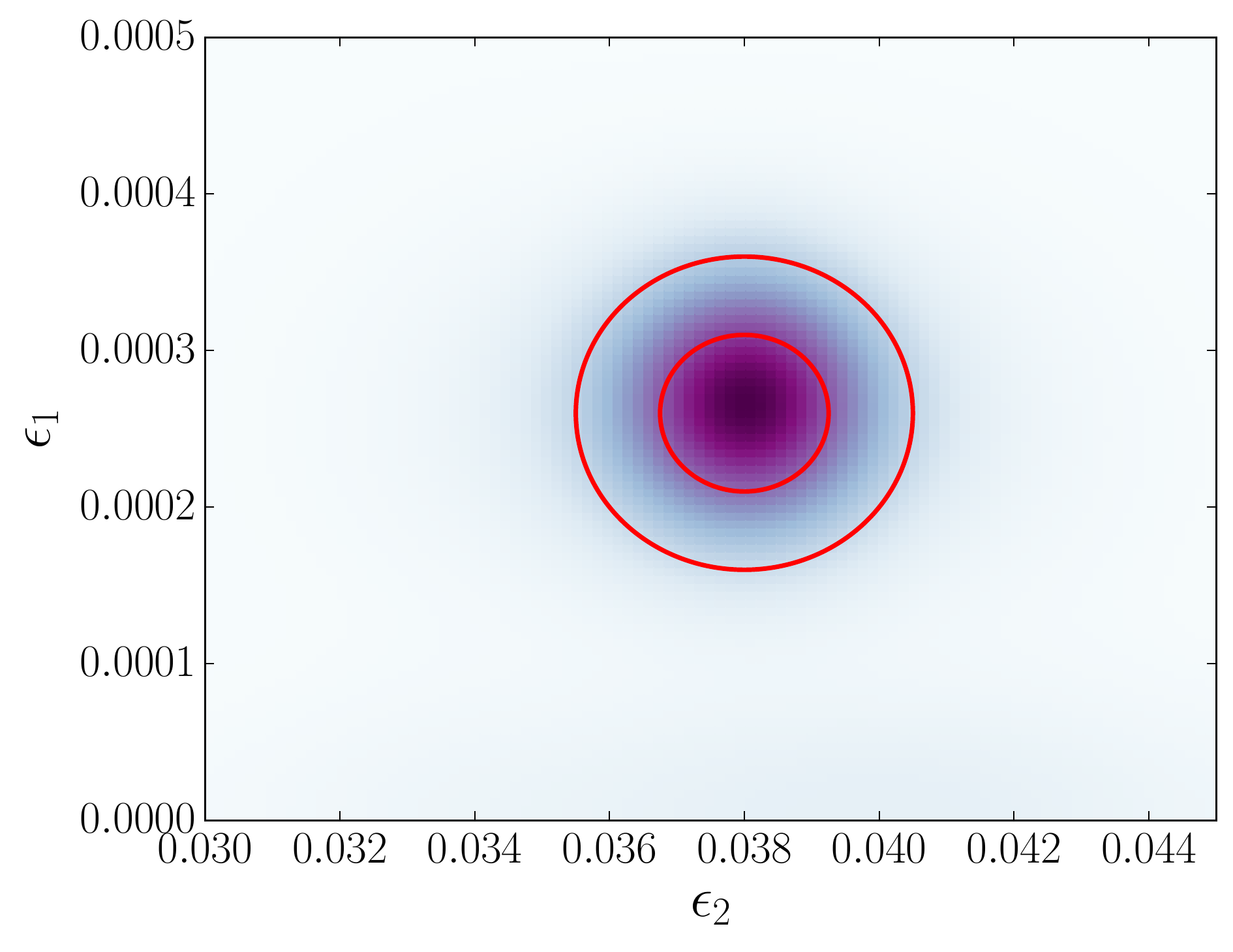}
\caption{~\label{fig:plot_eps1_eps2} A probability density plot indicating the shape of the LiteCOrE forecast likelihood (in purple) over a $(\epsilon_1,\epsilon_2)$ surface marginalised from the full $(\epsilon_1,\epsilon_2,\epsilon_3)$ space, illustrating the comparison with our Gaussian likelihood (the red contours). }
\end{center}
\end{figure}
%



When comparing the values from \texttt{MultiNest} and our method, we note that the former method is permitted many more samples from the model (in order to converge the integral for the Bayesian evidence) than the latter (which must limit the number of samples because many more computations of the same integral are required). Hence, the disagreement in values between the two methods that is not limited by the Gaussian likelihood assumption itself is likely to originate from this limitation of our computational resources.

The differences between the uncorrelated Gaussian likelihood and the sampled likelihood forecast for LiteCOrE (using the $\log \epsilon_1$ prior) are minute in the slicing of $(\epsilon_1,\epsilon_2)$-space depicted by \Fig{fig:plot_eps1_eps2}. Therefore, inaccuracies that can appear in the Bayesian evidence that arise from an imprecise analogy between a more realistic likelihood forecast and our mock forecasts are clearly far smaller than the disagreement that comes from our limited computational resources. The points for which the methods are in most disagreement are Category B and D (see appendix~\ref{sec:foxi-computation}), since they are characterised by a poor inter-point distance, but these points are sampled only very occasionally (see Table \ref{tab:percentage-points-categories}) and so we can expect minimal impact on our main conclusions in this work.

We shall leave the future application of our formalism to a proposed survey, such as COrE~\cite{Finelli:2016cyd}, for later work.

\section{Identifying the constraint on $\alphaS$}
\label{sec:alphaS-calculation}

To leading-order in the slow-roll expansion, the running of the scalar spectral index $\alphaS$ can be written for single-field models as
\begin{equation}
\alphaS \equiv \left. \frac{\dd^2 \ln {\cal P}_\zeta }{\dd (\ln k)^2} \right\vert_{k_*} \simeq - 2\epsilon_1 \epsilon_2 - \epsilon_2 \epsilon_3 \,.
\end{equation}
When no cross-correlations are observed --- as is the assumption in all of the forecast constraints in this work --- it can be shown that the generic cross-correlator from such a measurement reduces down to factors of correlators
\begin{align} \label{eq:alphaSection-cross-correlators}
\langle \epsilon_1^l \epsilon_2^m \epsilon_3^n \rangle &= \langle \epsilon_1^l\rangle \langle \epsilon_2^m\rangle \langle \epsilon_3^n\rangle\,.
\end{align}
For a Gaussian measurement on each of the slow-roll parameters, the fiducial point $\muf^{\alphaS} = \muf^{\alphaS}(\bmuf , \boldsigma ) \equiv \langle \alphaS \rangle$ can be derived from
\begin{align}
\muf^{\alphaS} &\simeq - 2\langle \epsilon_1 \epsilon_2 \rangle - \langle \epsilon_2 \epsilon_3 \rangle \,, \\
&\simeq  - 2\muf^1 \muf^2 - \muf^2 \muf^3 \label{eq:alphaS-fiducial-point-formula} \,.
\end{align}
The width of the measurement over $\alphaS$ can thus be unpacked into an expression containing only the fiducial points and widths on the slow-roll parameters, i.e. $\sigma^{\alphaS} = \sigma^{\alphaS}(\bmuf , \boldsigma )$
\begin{align}
\ (\sigma^{\alphaS})^2 &\equiv \langle \alphaS^2\rangle - \langle \alphaS \rangle^2 \nonumber \\
&\simeq \left\langle (2\epsilon_1 \epsilon_2 + \epsilon_2 \epsilon_3)^2\right\rangle - \left(  2\langle \epsilon_1 \epsilon_2 \rangle + \langle \epsilon_2 \epsilon_3 \rangle \right)^2 \nonumber  \\
&\simeq 4\langle \epsilon_1^2\rangle \langle \epsilon_2^2\rangle + \langle \epsilon_2^2\rangle \langle \epsilon_3^2\rangle + 4 \langle \epsilon_1\rangle \langle \epsilon_2^2\rangle  \langle \epsilon_3\rangle - 4\langle \epsilon_1\rangle^2 \langle \epsilon_2\rangle^2 - \langle \epsilon_2\rangle^2 \langle \epsilon_3\rangle^2 - 4 \langle \epsilon_1\rangle \langle \epsilon_2\rangle^2  \langle \epsilon_3\rangle \nonumber \\
&\simeq 4 (\sigma^1 )^2(\sigma^2 )^2 + 4 (\muf^1)^2(\sigma^2)^2 + 4 (\muf^2)^2(\sigma^1)^2 + (\sigma^2 )^2(\sigma^3 )^2 \nonumber \\
& \qquad \qquad \qquad + (\muf^2)^2(\sigma^3)^2 + (\muf^3)^2(\sigma^2)^2 + 4 \muf^1 \muf^3 (\sigma^2)^2 \label{eq:alphaS-width-formula}\,.
\end{align}
Using \Eq{eq:alphaS-fiducial-point-formula} and \Eq{eq:alphaS-width-formula} for a specified collection of widths on $\epsilon_1$, $\epsilon_2$ and $\epsilon_3$, we may identify all of the remaining fiducial points $\bmuf$ that satisfy a $2$-$\sigma$ measurement of $\alphaS$ and can therefore compute the probability defined in \Eq{eq:alphaS-probability}.

\clearpage
\bibliographystyle{JHEP}
\bibliography{decisiveness}

\end{document}